\documentclass{pasj01}
\draft 
\Received{$\langle$reception date$\rangle$}
\Accepted{$\langle$acception date$\rangle$}
\Published{$\langle$publication date$\rangle$}
\SetRunningHead{Astronomical Society of Japan}{Usage of \texttt{pasj00.cls}}
\usepackage{bm}
\usepackage{comment}
\usepackage{color}
\usepackage{url}
\usepackage{threeparttable}

\begin{document}

\title{Magnetic Field of Solar Dark Filaments Obtained from He I 10830 $\mathrm{\AA}$ Spectro-polarimetric Observation}
\author{Daiki \textsc{Yamasaki},\altaffilmark{1,}$^{*}$ Yu Wei \textsc{Huang},\altaffilmark{1} Yuki \textsc{Hashimoto}, \altaffilmark{1} Denis P. \textsc{Cabezas}, \altaffilmark{1} Tomoko \textsc{Kawate}, \altaffilmark{2,}\altaffilmark{3} Satoru \textsc{UeNo},\altaffilmark{1} and Kiyoshi \textsc{Ichimoto}\altaffilmark{1}}
\altaffiltext{1}{Astronomical Observatory, Kyoto University, Kitashirakawaoiwake-cho, Kyoto 606-8502, Japan}
\altaffiltext{2}{National Institute for Fusion Science, 322-6 Oroshi-cho, Toki, Gifu 509-5292 Japan}
\altaffiltext{3}{The Graduate University for Advanced Studies (SOKENDAI), Shonan-Kokusai Village, Hayama, Miura, Kanagawa 240-0193 Japan}
\email{dyamasaki@kusastro.kyoto-u.ac.jp}


\maketitle

\begin{abstract}
  Solar filaments are dense and cool plasma clouds in the solar corona.
  They are supposed to be supported in a dip of coronal magnetic field.
  However, the models are still under argument between two types of the field configuration; one is the normal polarity model proposed by Kippenhahn \& Schlueter (1957), and the other is the reverse polarity model proposed by Kuperus \& Raadu (1974).
  To understand the mechanism that the filaments become unstable before the eruption, it is critical to know the magnetic structure of solar filaments.
  In this study, we performed the spectro-polarimetric observation in the He I ($10830$ $\mathrm{\AA}$) line to investigate the magnetic field configuration of dark filaments.
  The observation was carried out with the Domeless Solar Telescope at Hida Observatory with a polarization sensitivity of $3.0\times10^{-4}$.
  We obtained 8 samples of filaments in quiet region.
  As a result of the analysis of full Stokes profiles of filaments, we found that the field strengths were estimated as $8$ - $35$ $\mathrm{G}$.
  By comparing the direction of the magnetic field in filaments and the global distribution of the photospheric magnetic field, we determined the magnetic field configuration of the filaments, and we concluded that 1 out of 8 samples have normal polarity configuration, and 7 out of 8 have reverse polarity configuration. 
\end{abstract}

\section{Introduction}
\begin{figure}[htb]
  \begin{center}
    \includegraphics[bb= 0 0 1380 1080, width=90mm]{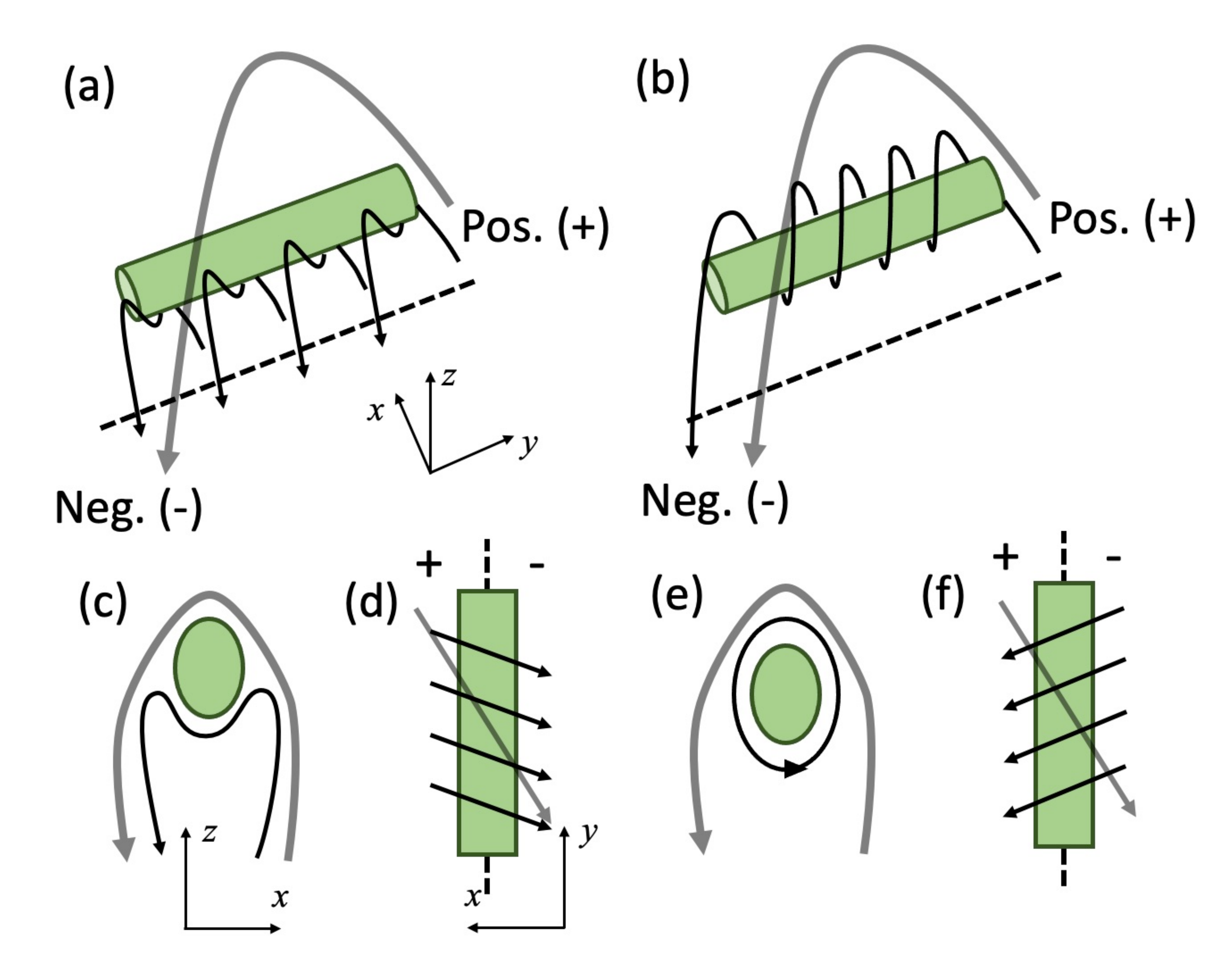}
  \end{center}
  \caption{Cartoons of two types of prominence models. Green cylinder, black solid lines, gray line, black bashed lines represent the prominence plasmas, magnetic field lines of a filament, a global magnetic field, and polarity inversion line, respectively. (a) Normal polarity model in 3-dimensional view, (b) Reverse polarity model in 3D view, (c) cross section in x-z plane of normal polarity model, (d) cross section in x-y plane of normal polarity model, (e) cross section in x-z plane of reverse polarity model, (f) cross section in x-y plane of reverse polarity model. Here z-axis is the normal to the solar surface and y-axis is along the magnetic neutral line.}\label{fig1}
\end{figure}

Solar dark filaments (or prominences seen on the solar limb) are cool ($\sim10^{4}$ $\mathrm{K}$) and dense ($>10^{9}$ $\mathrm{cm^{-3}}$) plasma suspended in the hot ($>10^{6}$ $\mathrm{K}$) corona.
They are supported by a bundle of twisted coronal magnetic field or magnetic flux ropes (MFRs: \cite{Xu2012}, \cite{Gibson2018}). 
Some of the dark filaments are located in quiet region, and some are in active region.
The former filaments are often called as quiescent (QS) filaments, and the latter are called as active region (AR) filaments.
Both of the QS and AR filaments locate above the polarity inversion lines (PILs: \cite{Babcock1955}).
The filaments sometimes become unstable and erupt into the interplanetary space, and solar flares are simultaneously observed \citep{Priest2002}.
A unified model of filament eruptions and solar flares are reviewed in \citet{Shibata2011}.
The eruptive solar flares are sometimes accompanied by the coronal mass ejections (CMEs), and the filaments are observed as a core structure of the CMEs \citep{Parenti2014}.
\\
~ To understand the formation process and the evolution of MFRs in ARs, a number of authors investigated the three-dimentional (3D) coronal magnetic field by using an extrapolation of the nonlinear force-free field (NLFFF: \cite{Inoue2013}, \cite{Kang2016}, \cite{Kawabata2017}, \cite{Muhamad2018},\cite{Yamasaki2021}).
In some cases, the NLFFF are used for initial conditions to investigate the dynamics of the erupting MFRs with a magnetohydrodynamic simulation (\cite{Jiang2013}, \cite{Inoue2015}, \cite{Inoue2018}, \cite{Inoue2021}, \cite{Yamasaki2022b}). 
However, the NLFFF approximation is not necessarily valid for the lower layer of the solar atmosphere such as the photosphere or low chromosphere.
This is because the magnetic pressure is not sufficiently higher than the gas pressure at these heights \citep{Gary2001}, $i.e.$, the assumption of the force-free is not valid.
\citet{Kawabata2020} reported that the shear angle ($i.e.$, angle from the potential field) of the horizontal magnetic field at the height of chromosphere reproduced with the NLFFF was smaller than that determined from observations.
\citet{Chaouche2012} performed two NLFFF extrapolations using observed photospheric and chromospheric magnetic field in dark filament regions independently, and by comparing the results of these two extrapolations they found that the difference in magnetic field structure that supports the filament plasmas; they found highly twisted field lines with more than one turn in the extrapolation from photosphere, but they only found less twisted field lines in the extrapolation from chromosphere.
This result suggests that the NLFFF only using a photospheric magnetogram is not suitable for reconstructing filaments' magnetic field.
Thus, direct observations of the magnetic field of the dark filaments are important to understand the structure and the mechanism of instability of the dark filaments. 
\\
~ In Table \ref{tab1}, we summarized the previous results on the magnetic field of dark filaments and prominences from spectro-polarimetric observations.
Regarding QS filaments, most of the studies suggest that the field strength is of the order of $10$ $\mathrm{Gauss}$ (\cite{Leroy1983}, \cite{Leroy1984}, \cite{Bommier1986}, \cite{Bommier1994}, \cite{Casini2003}, \cite{Orozco2014}, and \cite{Martinez2015}).
\citet{Wang2020} reported the field strength of an erupting QS filament was smaller than $100$ $\mathrm{Gauss}$.
Regarding AR filaments, some of the previous studies suggest that the field strength is $100-800$ $\mathrm{Gauss}$ (\cite{Xu2012}, \cite{Kuckein2009}, \cite{Sasso2011}, and \cite{Sasso2014}).
However, \citet{DiazBaso2016} pointed out that the observed Stokes profiles from the AR filaments can be explained with a field strength of $10$ $\mathrm{Gauss}$ in the filament if contaminations of observed spectra by the radiation from the background active region with 600 Gauss.
\\
~ Besides the field strength of QS and AR filaments, two types of the magnetic field configuration of the filaments are proposed. 
One is the normal polarity model \citep{Kippenhahn1957}.
In this model, the shear direction of the magnetic field of the filaments is the same as the shear direction of the ambient magnetic field inferred from the photospheric magnetic field (see Figure \ref{fig1} (a), (c), and (d)).
The other model is the reverse polarity model \citep{Kuperus1974}.
In this model, the shear direction of the magnetic field of the filaments is opposite to the shear direction of the ambient magnetic field (see Figure \ref{fig1} (b), (d), and (e)).
For QS filaments, \citet{Bommier1998} performed a spectro-polarimetric observation of an off-limb prominence in He I D3 line and found that the magnetic field structure was reverse polarity (see also \cite{Bommier1994}).
\citet{Martinez2015} performed a spectro-polarimetric observation of an off-limb prominence in He I $10830$ $\mathrm{\AA}$, and they also reported that the structure was the reverse polarity.
\citet{Wang2020} obtained the vector magnetic field of an erupting filament and reported that the magnetic field structure was reverse polarity during the eruption process.
For AR filaments, \citet{Xu2012} suggested that the magnetic field structure was normal polarity by using a simultaneous observation of He I $10830$ $\mathrm{\AA}$ and Si I $10827$ $\mathrm{\AA}$ lines.
\citet{Sasso2014} analyzed the vector magnetic field of filament and concluded that the magnetic field topology was reverse polarity configuration.
\citet{Yokoyama2019} studied the photospheric vector magnetic field of a filament channel in an AR and suggested that the field configuration was reverse polarity. 
\\
~ Diagnostics of the magnetic field structure of the solar filaments require vector magnetic field information. 
Spectro-polarimetric observation of the absorption line of the He I $10830$ $\mathrm{\AA}$ is one of the most powerful methods to quantitatively obtain the magnetic field of the solar filaments \citep{Hanaoka2017}.
This line is sensitive to magnetic field not only through the Zeeman effect \citep{Zeeman1897} but also through the Hanle effect \citep{Hanle1924}.
From the Zeeman effect in this line, the longitudinal components of $10$ $\mathrm{Gauss}$ and transverse components of $200$ $\mathrm{Gauss}$ are detectable at a sensitivity of $0.03\%$ in polarimetric measurements.
In contrast, the Hanle effect diagnostics allow us to deduce relatively weak field of $0.1-100$ $\mathrm{Gauss}$.
In this study, to clarify not only the magnetic field strength of the filaments but also the magnetic field configuration, we performed a spectro-polarimetirc observation in He I $10830$ $\mathrm{\AA}$ line targeting dark filaments located in quiet regions.
\\
~ The rest of this paper is structured as follows; the observation is introduced in Section \ref{sec:obs}, the analysis is described in Section \ref{sec:ana}, results are presented in Section \ref{sec:res}, and discussion and conclusion on our findings are given in Section \ref{sec:dis}.

\begin{threeparttable}[htb]
  \caption{Past observational works on filament(/prominence) magnetic field}
  \begin{center}
    \label{tab1}
    \begin{tabular}{lcccc}
      \hline
      Paper                 & Filament  & Field strength & Normal pol.        & Reverse pol.        \\
                            & type      & [Gauss]        & (K-S model)        & (K-R model)         \\
      \hline
      \hline
      \citet{Wang2020}      & QS        & $<100$         & -                  & $\circ$             \\
      \citet{Casini2003}    & QS        & $10-70$        & -                  & -                   \\
      \citet{Orozco2014}    & QS        & $\sim25$       & -                  & -                   \\
      \citet{Martinez2015}  & QS        & $<20$          & -                  & $\circ$             \\
      \citet{Bommier1986}   & QS        & $2-20$         & -                  & -                   \\
      \citet{Leroy1984}     & QS        & $5-10$         & -                  & -                   \\
      \citet{Bommier1994}   & QS        & $7.5$          & -                  & $\circ$             \\
      \citet{Leroy1983}     & QS        & $6$ \tnote{1}  & -                  & -                   \\
      \citet{Bommier1998}   & QS        & -              & -                  & $\circ$             \\
      \hline
      \citet{Xu2012}        & AR        & $600-800$      & $\circ$            & -                   \\
      \citet{Kuckein2009}   & AR        & $<500$         & -                  & -                   \\
      \citet{Sasso2011}     & AR        & $100-250$      & -                  & -                   \\
      \citet{Sasso2014}     & AR        & $\sim100$      & -                  & $\circ$             \\      
      \citet{DiazBaso2016}  & AR        & $\sim10$       & -                  & -                   \\
      \citet{Yokoyama2019}  & AR        & -              & -                  & $\circ$ \tnote{2}   \\
      \hline
    \end{tabular}
    \begin{tablenotes}
      \item[1] Their target was a polar crown.
      \item[2] By photospheric magnetic field observation below dark filaments.
    \end{tablenotes}
  \end{center}
\end{threeparttable}

\section{Observation} \label{sec:obs}
\begin{figure}[htb]
  \begin{center}
    \includegraphics[bb= 0 0 1400 670, width=160mm]{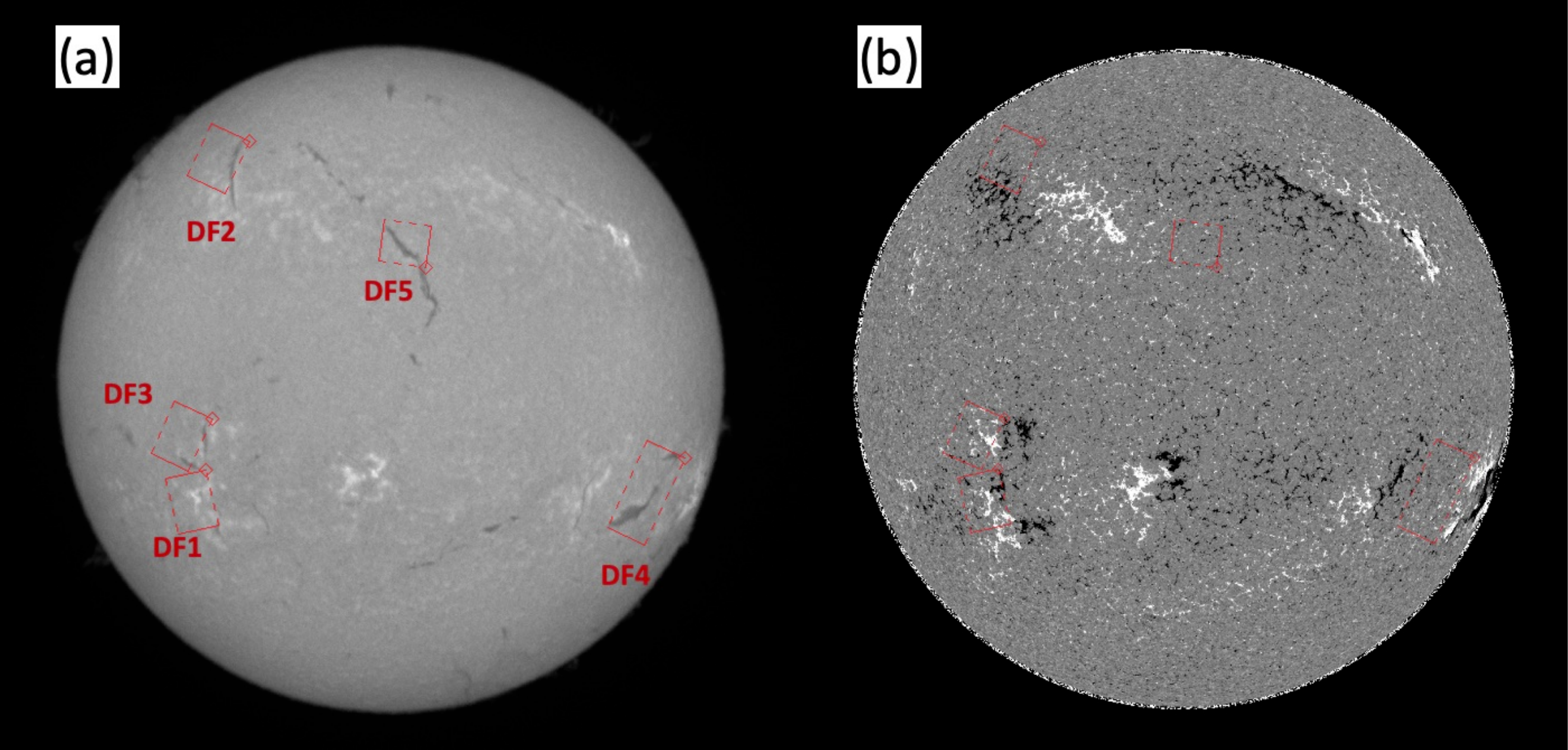}
  \end{center}
  \caption{Location of the observational targets taken on 2022 Apr 9. (a) H$\alpha$ line center image taken by the SMART-SDDI. (b) Photospheric magnetogram taken with SDO-HMI. Greyscale show the radial component of the magnetic field. White and black correspond to the field strength of $-50$ and $50$ Gauss, respectively. Red line rectangles show the field-of-view of the target DF regions. Solid and dashed lines represent slit and scan directions, respectively. Diamond symbol represents the origion of the field of view.}\label{fig21}
\end{figure}

\begin{figure}[htb]
  \begin{center}
    \includegraphics[bb= 0 0 1400 670, width=160mm]{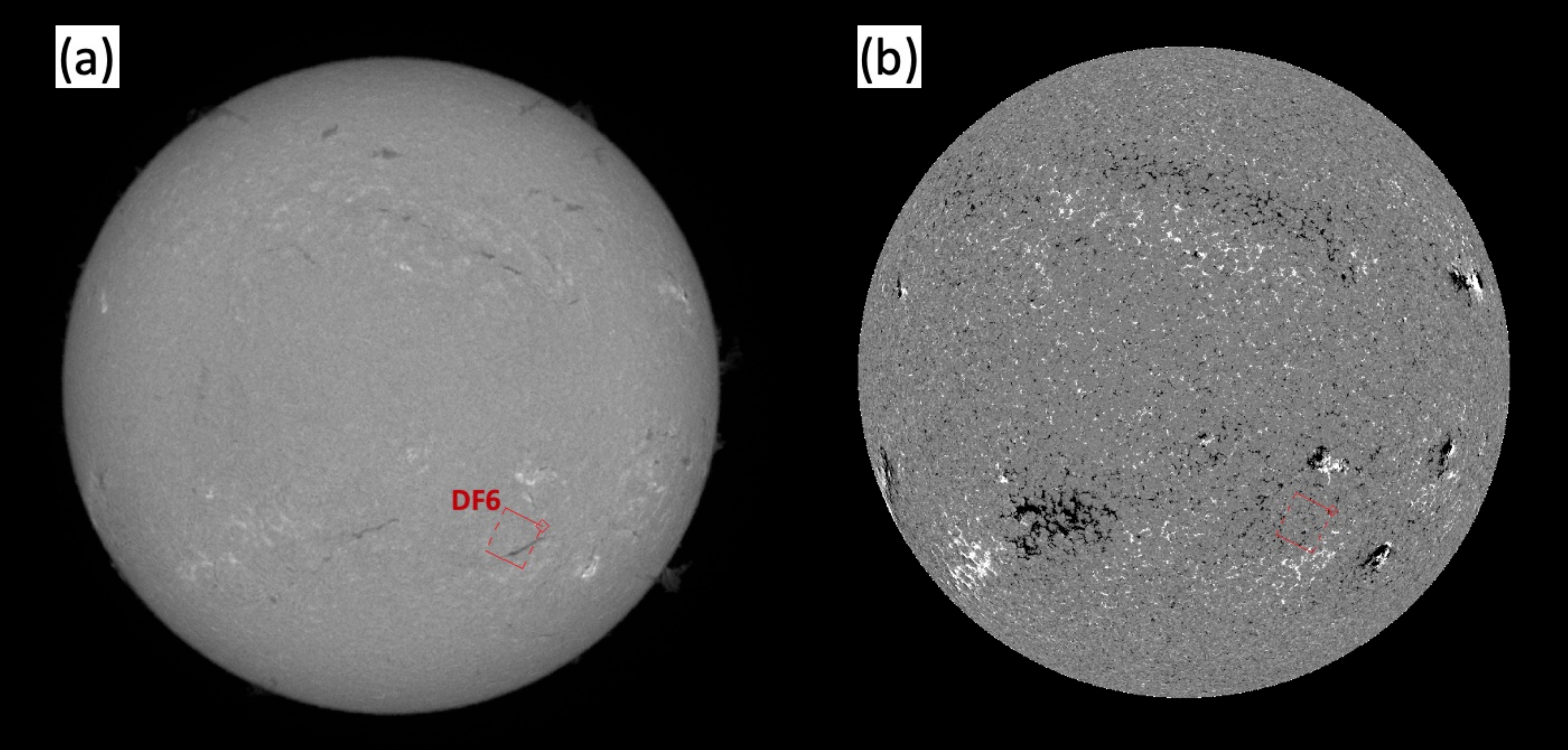}
  \end{center}
  \caption{Location of the observational targets taken on 2022 Jun 4. Format of this figure is same as that of Figure \ref{fig21}.}\label{fig22}
\end{figure}

\begin{figure}[htb]
  \begin{center}
    \includegraphics[bb= 0 0 1400 670, width=160mm]{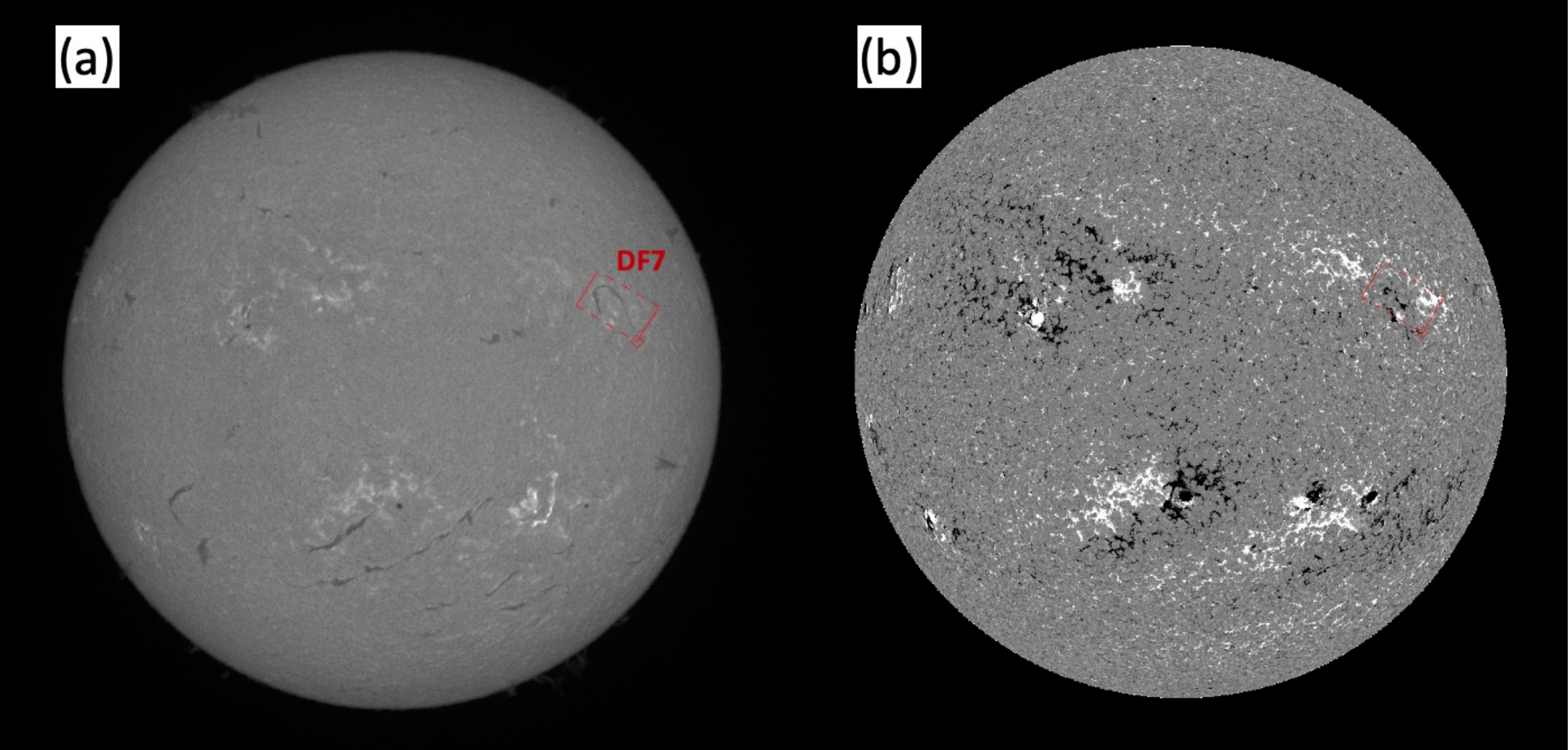}
  \end{center}
  \caption{Location of the observational targets taken on 2022 Aug 11. Format of this figure is same as that of Figure \ref{fig21}.}\label{fig23}
\end{figure}

\begin{figure}[htb]
  \begin{center}
    \includegraphics[bb= 0 0 1400 670, width=160mm]{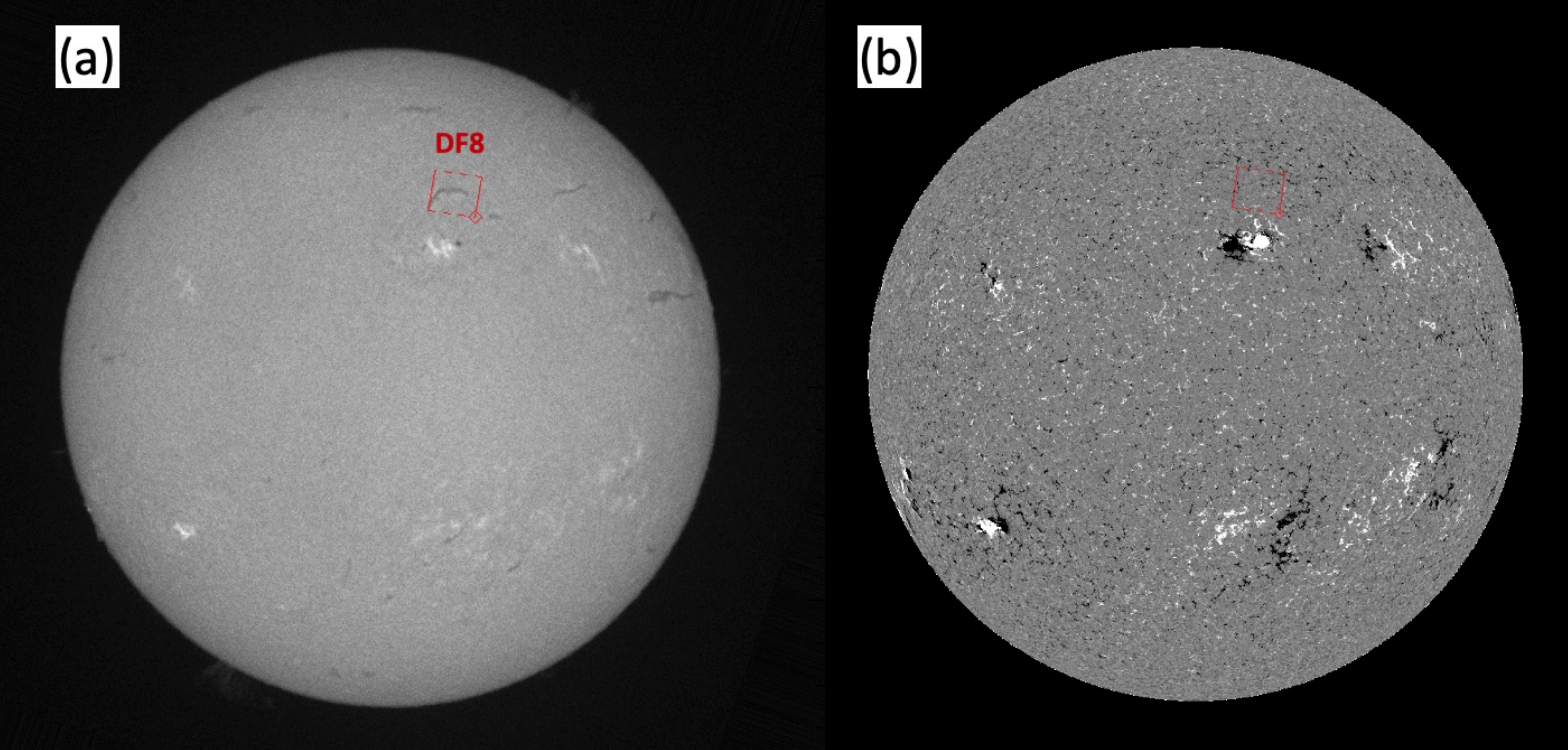}
  \end{center}
  \caption{Location of the observational targets taken on 2022 Aug 24. Format of this figure is same as that of Figure \ref{fig21}.}\label{fig24}
\end{figure}

Observation was performed by using the newly installed spectro-polarimeter on the Domeless Solar Telescope (DST: \cite{Nakai1985}) at Hida Observatory \citep{Ichimoto2022}.
The polarimeter consisting of a rotating super achromatic waveplate and a polarizing beam splitter is located just behind the slit of the spectrograph, and it provides efficient modulations for two orthogonally polarized spectra on detectors in wavelength range of $5000-11000$ $\mathrm{\AA}$.
In our observation in He I $10830$ $\mathrm{\AA}$, polarimetric sensitivity of $3\times10^{-4}$ was achieved in a few second with a near-infrared (NIR) camera, the Goldeye G033 SWIR ($640\times512$ pixels), (see \cite{Yamasaki2022a}).
\\
~ Using the spectro-polarimeter, we performed the observations of 8 dark filaments (DFs) on 2022 Apr 9, 2022 Jun 4, 2022 Aug 11, and 2022 Aug 24 JST (see Table \ref{tab3}).
Hereafter, we use filament IDs shown in Table \ref{tab3} to identify each of the target filaments.
He I $10830$ $\mathrm{\AA}$ and Si I $10827$ $\mathrm{\AA}$ lines were taken with the NIR camera with an exposure time of $15$ $\mathrm{msec}$ for each frame.
We obtained $200$ frames in $3$ $\mathrm{sec}$ for each slit position with $1.0$ $\mathrm{Hz}$ modulation by the rotating waveplate. 
The slit width and length were $0.1$ and $20$ $\mathrm{mm}$, corresponding to $0''.64$ and $128''$ on the solar image, respectively.
Spatial sampling along the slit is $0''.43$ $\mathrm{pixel^{-1}}$, and the spectral sampling is $29$ $\mathrm{m\AA~ pixel^{-1}}$ while scan step was $1''.38$.
Fields-of-view along the slit direction ($X$) and the scan direction ($Y$) for each observation is given in Table \ref{tab3}.
\\
~ The 8 observational targets are indicated on the full disk H$\alpha$ solar images taken by the Solar Dynamics Doppler Imager (SDDI: \cite{Ichimoto2017}) on the Solar Magnetic Activity Research Telescope (SMART: \cite{UeNo2004}) in Figures \ref{fig21}, \ref{fig22}, \ref{fig23}, and \ref{fig24} (a).
In addition, we also show full disk images of the photospheric magnetic field taken by the Helioseismic and Magnetic Imager (HMI: \cite{Scherrer2012}) onboard the $Solar~ Dynamics~ Observatory$ ($SDO$: \cite{Pesnell2012}) in Figures \ref{fig21}, \ref{fig22}, \ref{fig23}, and \ref{fig24} (b).
From the photospheric magnetogram and the H$\alpha$ images, we identified that the DF1, DF3, and DF7 locate in plage regions, the DF2, DF4, DF5, and DF6 locate in quiet regions, and the DF8 locate near an active region.
\\
~ Geometrical height of the target filaments is an important parameter for a correct evaluation of the magnetic field by using the Hanle effect.
In our study, we estimated the heights of the targets using the SMART-SDDI images days prior or posterior to the DST spectro-polarimetric observation at which they were off-limb prominences.
The adopted heights for each filament are shown in Table \ref{tab3}.

\begin{threeparttable}[htb]
  \caption{Observational targets}
  \begin{center}
    \label{tab3}
    \begin{tabular}{lccrrrrrr}
      \hline
      ID           & Observation  & Scan time in UT   &           &  Position \tnote{1} &                  & $X$ \tnote{2} & $Y$ \tnote{2} & Height  \\
                   & date         & (start-end)       & $r$       &  $p$                &  $i$             & [$''$]        & [$''$]        & [$''$]  \\
      \hline
      DF1 \tnote{3} & 2022 Apr 8   & 23:34:41-23:42:36 & 11' 16'' &  95$^{\circ}$ 23'    &  76$^{\circ}$ 30' & 128 & 166 & 25 \\
      DF2           & 2022 Apr 9   & 00:37:28-00:45:23 & 13' 31'' &  11$^{\circ}$ 47'    &  39$^{\circ}$ 36' & 128 & 166 & 17 \\
      DF3 \tnote{3} & 2022 Apr 9   & 01:40:53-01:48:48 & 10' 28'' &  78$^{\circ}$ 58'    &  39$^{\circ}$ 25' & 128 & 166 & 21 \\
      DF4           & 2022 Apr 9   & 02:15:46-02:29:00 & 13' 27'' & 219$^{\circ}$ 48'    &  39$^{\circ}$ 10' & 128 & 276 & 23 \\
      DF5           & 2022 Apr 9   & 03:02:31-03:09:06 &  6' 35'' & 327$^{\circ}$ 57'    & 326$^{\circ}$ 04' & 128 & 138 & 29 \\
      DF6           & 2022 Jun 4   & 05:42:41-05:49:16 &  9' 34'' & 204$^{\circ}$ 30'    &  50$^{\circ}$ 16' & 128 & 138 & 18 \\
      DF7 \tnote{3} & 2022 Aug 11  & 06:25:28-06:35:22 & 11' 34'' & 302$^{\circ}$ 50'    & 343$^{\circ}$ 43' & 128 & 207 & 15 \\
      DF8 \tnote{4} & 2022 Aug 24  & 00:57:59-01:04:34 &  9' 30'' & 359$^{\circ}$ 27'    &  10$^{\circ}$ 41' & 128 & 138 & 19 \\
      \hline
    \end{tabular}
    \begin{tablenotes}
      \item[1] The position parameters of $r$, $p$, and $i$ represent the heliocentric distance, polar angle measured from the geographic north, and inclination angle of the slit of spectrograph from the geographical N-S direction, respectively.
      \item[2] $X$ is the field view along the slit common for all the targets and $Y$ is the scanning range.
      \item[3] Targets locate above plage regions (see Figure \ref{fig21} for DF1 and DF3 and Figure \ref{fig23} for DF8).   
      \item[4] One footpoint of the filament is connected to an active region (see Figure \ref{fig24}).
    \end{tablenotes}
  \end{center}
\end{threeparttable}

\section{Analysis}\label{sec:ana}
\subsection{Data reduction and calibration}
\begin{figure}[htb]
  \begin{center}
    \includegraphics[bb= 0 0 1270 560, width=160mm]{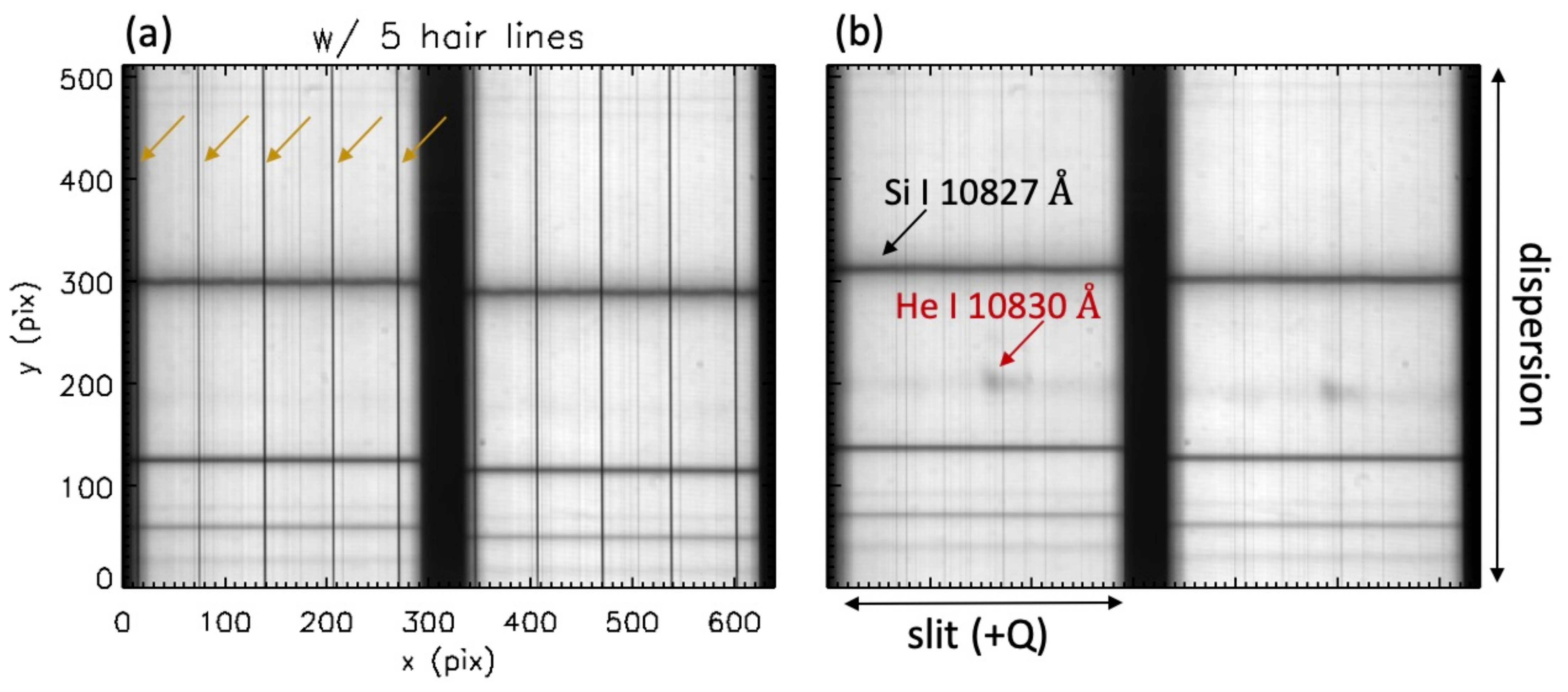}
  \end{center}
  \caption{(a) Example of a calibration data with 5 hair lines. Yellow arrows indicate the hair lines. (b) Example of an observed raw spectra. Black and red arrows indicate the Si I $10827$ $\mathrm{\AA}$ and the He I $10830$ $\mathrm{\AA}$ lines, respectively. Horizontal and vertical axes correspond to the slit and dispersion directions, respectively. The slit direction defines the $+Q$ axis.}\label{fig311}
\end{figure}

The obtained spectral data were reduced with dark frame subtraction and flat-fielding.
The orthogonally polarized spectra recorded simultaneously by the NIR camera were aligned after correcting the distortion of spectral images by referring to 5 hair lines imaged on the spectra taken after the observation sequence (see Figure \ref{fig311}).
Then we performed polarization demodulation for each set of 200 images, and combined them to obtain Stokes $I$, $Q$, $U$, and $V$ spectrum. 
Here we note that the slit direction defines the $+Q$ direction.
Regarding the calibration of the instrumental polarization of the DST, we referred to \citet{Anan2018}.
For an initial guess of the M\"{u}ller matrix of the DST, we used the data obtained by \citet{Anan2012}.
Further adjustment of the matrix is done by making the Zeeman signals in Si I $10827$ $\mathrm{\AA}$ line to be symmetric ($Q$ and $U$) or anti-symmetric ($V$) in a sunspot observed before or after the filament observation.
After polarization calibration, we removed artificial fringe pattern in Stokes $Q$, $U$, and $V$ spectra by subtracting periodic component in dispersion direction.
For more details of the calibration procedure, see \citet{Ichimoto2022}.

\subsection{Stokes Inversions}
\subsubsection{Coordinate system}
\begin{figure}[htb]
  \begin{center}
    \includegraphics[bb= 0 0 550 460, width=80mm]{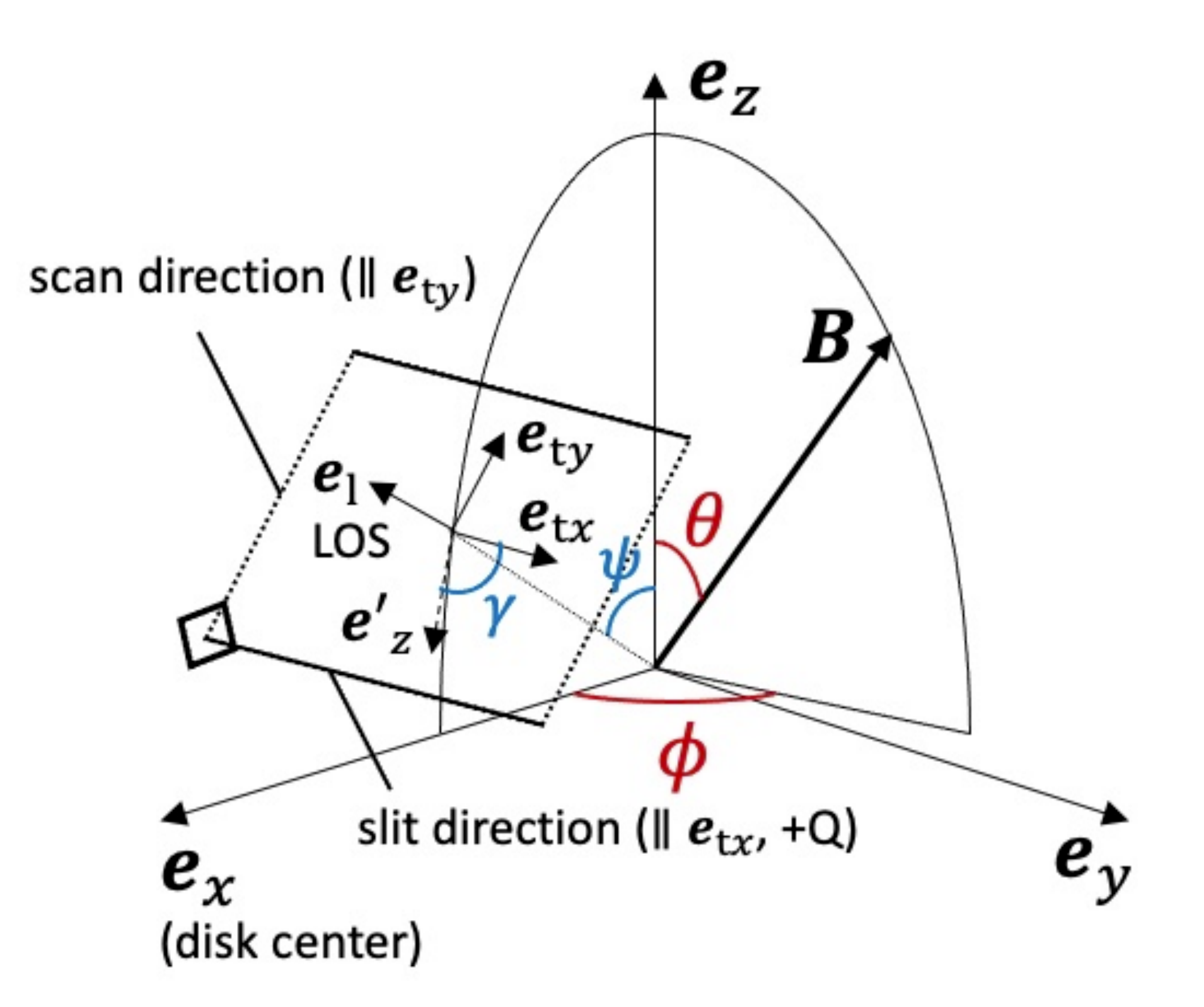}
  \end{center}
  \caption{Definition of the coordinate systems in the local frame and the observer's frame. The vectors $\bm{e_{x}}, \bm{e_{y}}$, and $\bm{e_{z}}$ correspond to the cartesian coordinate units in local frame. $\theta$ and $\phi$ represent the inclination and azimuth angles of the magnetic field vector, $\bm{B}$ with respect to the local frame coordinate system. The vectors $\bm{e_{\mathrm{t}x}}, \bm{e_{\mathrm{t}y}}$, and $\bm{e_{\mathrm{l}}}$ correspond to cartesian coordinate units in observers' frame where $\bm{e_{\mathrm{l}}}$ is in line-of-sight. $\gamma$ represents the angle between $\bm{e_{\mathrm{t}x}}$ and the direction of solar radius ($\bm{e'_{z}}$). $\psi$ is the angle between $\bm{e_{z}}$ and $\bm{e_{\mathrm{l}}}$. Diamond symbol represents the origin of the field of view. Solid and dashed lines correspond to the slit and scan directions, respectively.}\label{fig321}
\end{figure}

In our study, we use two different coordinate systems as given in \citet{AsensioRamos2008}: the local frame at the target region on the sun and the observer's frame.
Figure \ref{fig321} show these coordinate systems.
The vectors $\bm{e_{x,y,z}}$ are the unit vectors of the cartesian coordinate in local frame, and the vectors $\bm{e_{\mathrm{t}x, \mathrm{t}y, \mathrm{l}}}$ are the unit vectors of the cartesian coordinate in observers' frame where $\bm{e_{\mathrm{t}x}}$, $\bm{e_{\mathrm{t}y}}$, and $\bm{e_{\mathrm{l}}}$ represent the slit, scan, and line-of-sight directions, respectively.
We denote the vector magnetic field ($\bm{B}$) in the local frame as $B_{x,y,z}$.
By using the field strength ($|\bm{B}|$), the inclination ($\theta$) and the azimuth ($\phi$) of the magnetic field in the local coordinate, we can describe each of the magnetic field components as follows:
\begin{eqnarray}
  B_{x} &=& |\bm{B}|\sin\theta\cos\phi, \\
  B_{y} &=& |\bm{B}|\sin\theta\sin\phi, \\
  B_{z} &=& |\bm{B}|\cos\theta.
\end{eqnarray}
In this paper, we call $B_{x}$ and $B_{y}$ as the $horizontal$ $component$ and $B_{z}$ as the $vertical$ $component$ of the magnetic field, respectively.
We denote the magnetic field in the observer's frame as $B_{\mathrm{t}x, \mathrm{t}y, \mathrm{l}}$.
The relation between the magnetic field components in the local frame and the observer's frame is given by
\begin{eqnarray}
  B_{\mathrm{t}x} &=& B_{y}\sin\gamma + (B_{x}\cos\psi - B_{z}\sin\psi)\cos\gamma, \\
  B_{\mathrm{t}y} &=& B_{y}\cos\gamma - (B_{x}\cos\psi - B_{z}\sin\psi)\sin\gamma, \\
  B_{\mathrm{l}}  &=& B_{z}\cos\psi + B_{x}\sin\psi,
\end{eqnarray}
where $\gamma$ is the angle between $\bm{e_{\mathrm{t}x}}$ and the direction of solar radius ($\bm{e'_{z}}$), and $\psi$ is the angle between $\bm{e_{z}}$ and $\bm{e_{\mathrm{l}}}$ (see Figure \ref{fig321}).
In this paper, we call $B_{\mathrm{t}x}$ and $B_{\mathrm{t}y}$ as the $transverse$ $component$ and $B_{\mathrm{l}}$ as the $longitudinal$ $component$, respectively.

\subsubsection{Stokes inversion}
To deduce the photospheric vector magnetic field, we applied the Milen Eddington inversion to the full Stokes profiles of Si I $10827$ $\mathrm{\AA}$ line.
The inversion code includes 8 physical parameters: the magnetic field strength, the inclination and azimuth of the magnetic field vector in observer's frame, the line of sight velocity, the turbulent velocity, the damping constant of the line, the continuum intensity, and the ratio of the opacities in line center and continuum.
\\
~ To obtain the magnetic field in the filaments, we performed the Stokes inversion for full Stokes profiles of He I $10830$ $\mathrm{\AA}$ by using the HAZEL code developed by \citet{AsensioRamos2008}.
This inversion code takes into account not only the Zeeman effect but also the atomic polarization.
The HAZEL obtains 8 physical parameters from the fitting of observed Stokes profiles, $i.e.,$ the magnetic field strength ($|\bm{B}|$), the inclination ($\theta$) and azimuth ($\phi$) of the magnetic field vector with respect to the local vertical, the optical thickness ($\tau$), the Doppler velocity ($v_{\mathrm{doppler}}$), the turbulent velocity ($v_{\mathrm{turb}}$), the line damping ($a$), and the filling factor ($ff$).
To prevent unexpected results due to the Van-Vleck ambiguity (\cite{LandiDegl'Innocenti2004}, \cite{AsensioRamos2008}), we performed the fitting for 3 different ranges of the inclination angle separately.
They are referred to ``case A'' for $0^{\circ}.0<\theta<54^{\circ}.74$, ``case B'' for $54^{\circ}.74<\theta<125^{\circ}.26$, and ``case C'' for $125^{\circ}.26<\theta<180^{\circ}.0$ in this paper. 
Regarding the other physical parameters, we set the range into $0.0<|\bm{B}|<500.0$ $\mathrm{Gauss}$, $-180^{\circ}.0<\phi<180^{\circ}.0$, $0.1<\tau<5.0$, $-20.0<v_{\mathrm{doppler}}<20.0$ $\mathrm{km/s}$, $3.0<v_{\mathrm{turb}}<15.0$ $\mathrm{km/s}$, $0.0<a<1.0$, and fixed $ff=1.0$.
Regarding the height of the dark filaments, we used the values shown in Table \ref{tab3}.

\section{Results}\label{sec:res}
\subsection{Stokes signals}
\begin{figure}[htb]
  \begin{center}
    \includegraphics[bb= 0 0 830 1050, width=120mm]{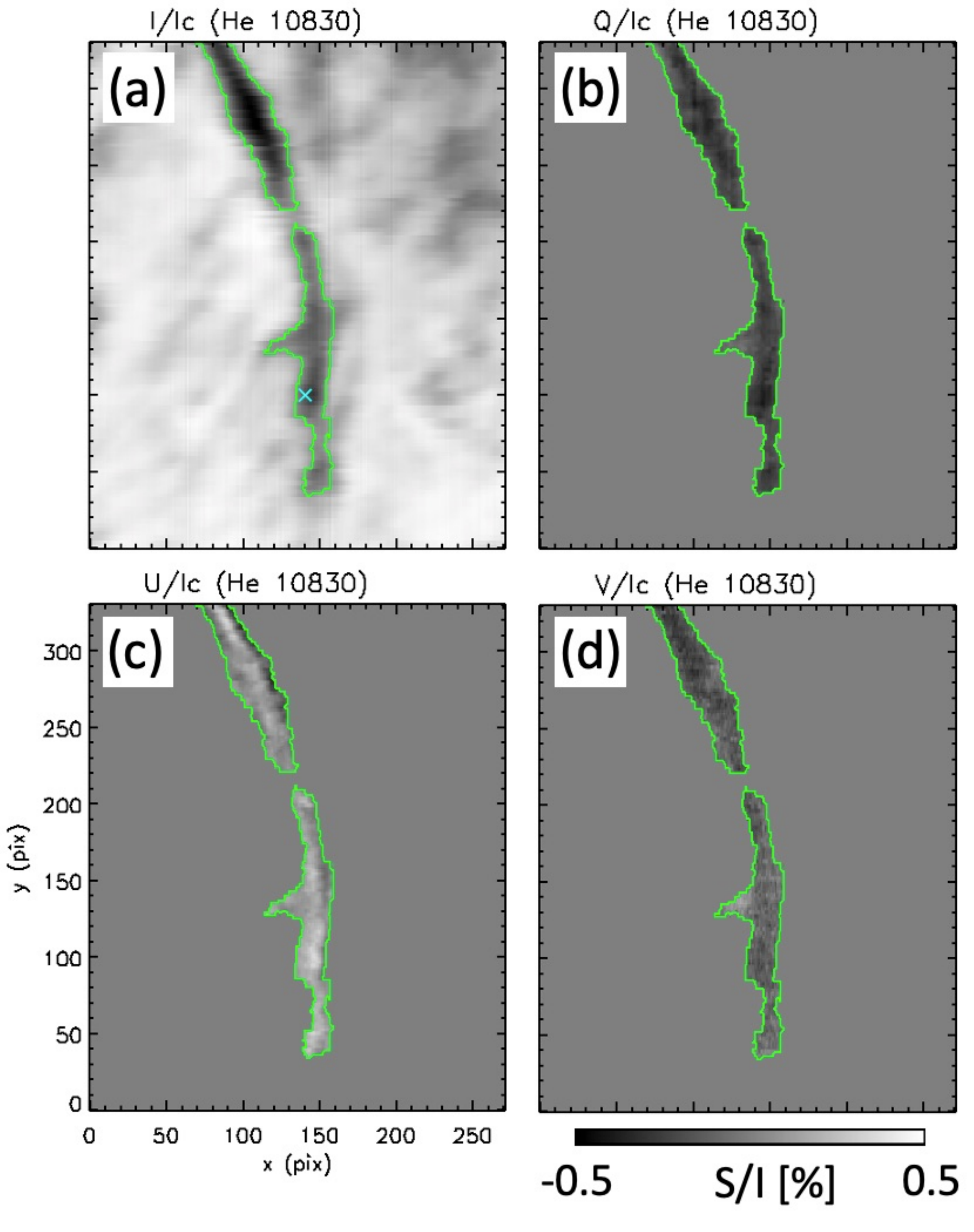}
  \end{center}
  \caption{Distribution of Stokes signal for DF2. (a) Stokes I image at the line center of He I 10830 $\mathrm{\AA}$. Blue cross symbol in the panel shows the location of the pixel of interest. (b,c) Stokes Q and U integrated values in a range of $\pm0.25$ $\mathrm{\AA}$ around the line center of He I 10830 $\mathrm{\AA}$ and normalized by continuum intensity. (d) Stokes V image made from subtraction of red ($0.25$ $\mathrm{\AA}$) and blue ($-0.25$ $\mathrm{\AA}$) wing normalized by continuum intensity. Green contour show the mask for the target dark filament region. }\label{fig411}
\end{figure}

\begin{figure}[htb]
  \begin{center}
    \includegraphics[bb= 0 0 920 700, width=150mm]{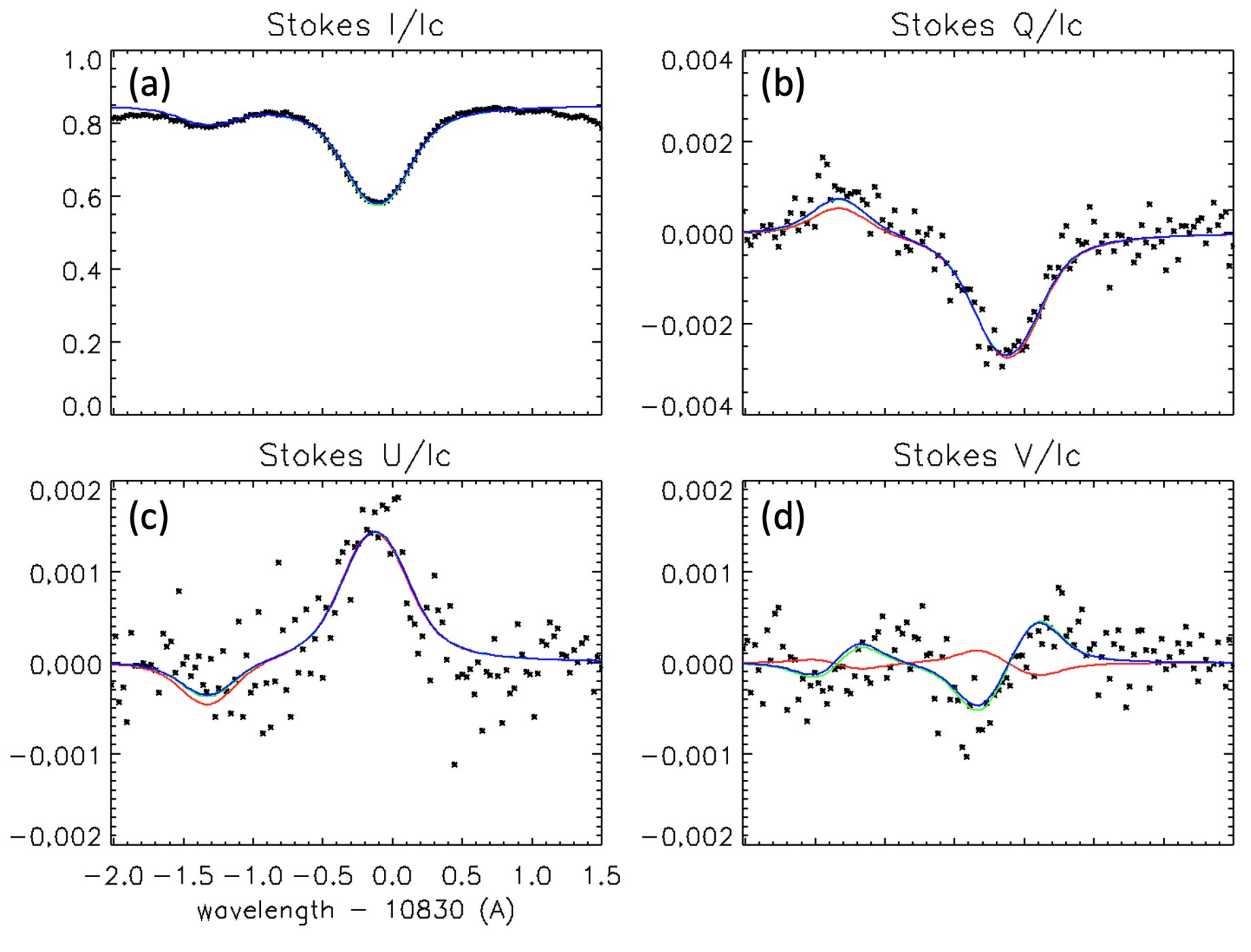}
  \end{center}
  \caption{Stokes profiles of I, Q, U, and V at pixel of interest of DF2 (see the red symbol in Figure \ref{fig411} (a)). Black symbols show the observational data. Red, green, and blue solid lines show the fitting results for ``case A'', ``case B'', ``case C'', respectively.}\label{fig412}
\end{figure}

In Figure \ref{fig411}, we show the distribution of the Stokes signals in DF2.
All the Stokes signals are normalized by the contiuum intensity.
Panel (a) displays the intensity at the line center of He I $10830$ $\mathrm{\AA}$, where we can identify the target filament as a dark structure.
The blue symbol in panel (a) indicates the location of the pixel for which Stokes profiles are shown in Figure \ref{fig412}. 
Green contour shows the border of the filament determined from the Stokes $I$ map and represents the mask for the target region.
Regarding the linear polarization signals of Stokes $Q$ and $U$, we display the integrated values in range of $\pm0.25$ $\mathrm{\AA}$ around the line center.
We find negative and positive signals of up to $\sim0.5\%$ through the filament body in Stokes $Q$ and $U$, respectively (see panels (b) and (c) in Figure \ref{fig411}).
Regarding the circular polarization signal of Stokes $V$, we display the value of the subtraction of the red wing at $+0.25$ $\mathrm{\AA}$ and blue wing at $-0.25$ $\mathrm{\AA}$ ($V_{\mathrm{blue}}-V_{\mathrm{red}}$).
We find negative signals through the filament body (see panel (d) in Figure \ref{fig411}). 
\\
~ Black symbols in Figure \ref{fig412} shows the observed Stokes profiles around He I $10830$ $\mathrm{\AA}$ at the cyan pixel shown in Figure \ref{fig411} (a).
In Figure \ref{fig412}, we find $-0.3\%$ of the Stokes $Q$ signal, $0.2\%$ of Stokes $U$ signal, and $\pm0.05\%$ of Stokes $V$ signal in observed profiles.
Regarding the linear polarization signals, red and blue components of the Stokes $Q$ and $U$ show the opposite sign, suggesting that these signals are caused by the Hanle effect \citep{TrujilloBueno2002}.
Regarding the circular polarization signal, both red and blue components show the anti-symmetric profiles, suggesting that this signal is caused by the longitudinal Zeeman effect.

\subsection{Inversion results} \label{subsec:42}
\begin{figure}[htb]
  \begin{center}
    \includegraphics[bb= 0 0 1025 895, width=150mm]{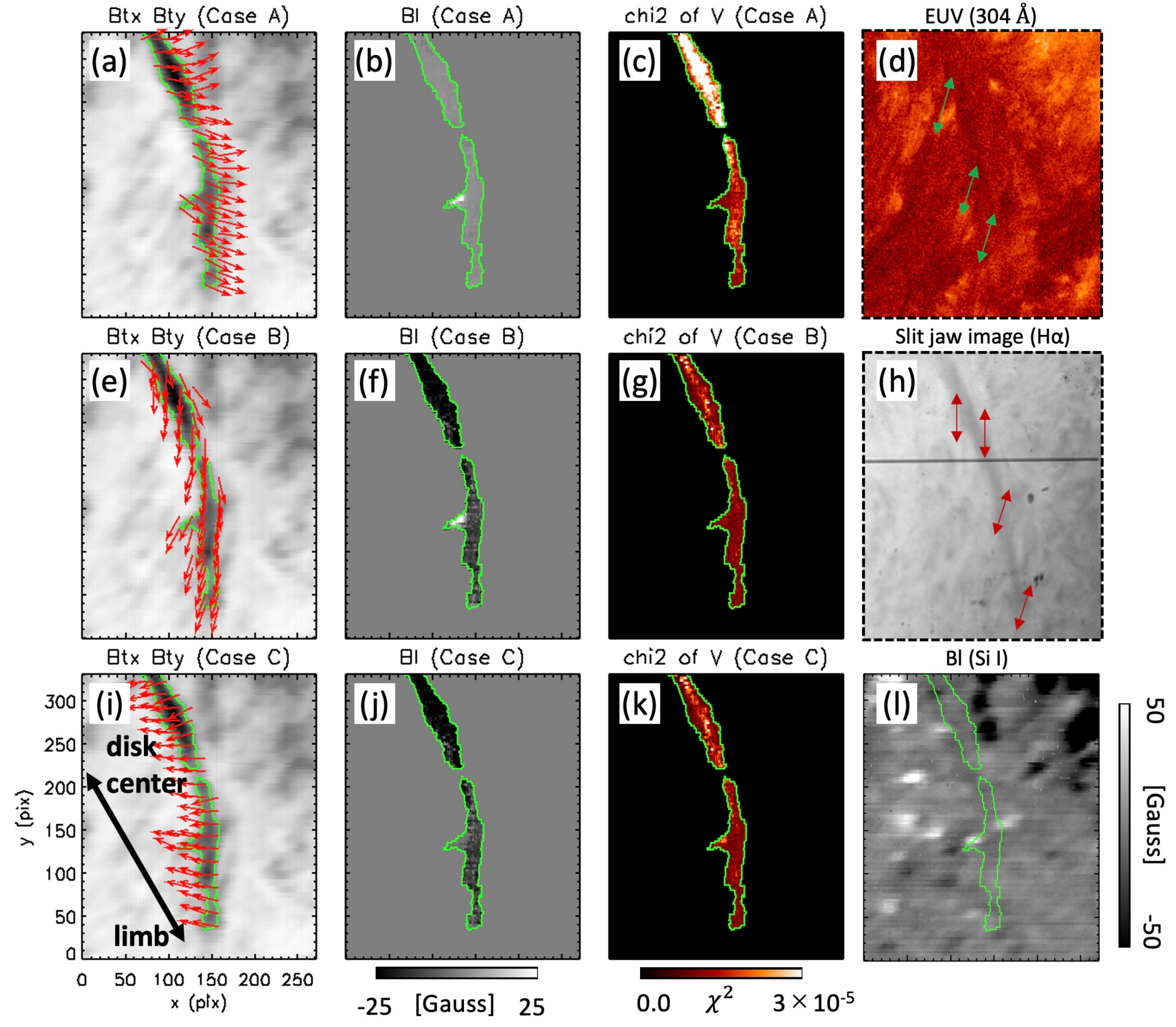}
  \end{center}
  \caption{Vector magnetic field for DF2 obtained from the HAZEL inversion. Green contour show the mask for the target dark filament region. (a) Transverse magnetic field vector in observer's frame with red arrows are overplotted on Stokes I image at the line center of He I 10830 $\mathrm{\AA}$ for ``Case A'' result. (b) Longitudinal magnetic field in observer's frame for ``Case A'' result. (c) $\chi^{2}$ distribution of Stokes $V$ for ``Case A'' result. (d) EUV $304$ $\mathrm{\AA}$ image taken with AIA. Green arrows show the direction of the threads of the filament. (e) Same as (a) for ``Case B'' result. (f) Same as (b) for ``Case B'' result. (g) Same as (c) for ``Case B'' result. (h) Slit jaw image at H$\alpha$ line center, 6563 $\mathrm{\AA}$. Red arrows show the direction of the threads of the filament. (i) Same as (a) for ``Case C'' result. (j) Same as (b) for ``Case C'' result. (k) Same as (c) for ``Case C'' result. (l) Line-of-sight component of the photospheric magnetic field obtained from Si I $10827$ $\mathrm{\AA}$ line observation. White- and black-colored regions correspond to positive and negative polarities, respectively.}\label{fig421}
\end{figure}

In Figure \ref{fig412}, red, green, and blue solid curves show the fitting results from the HAZEL inversion for the ``case A'', ``case B'', and ``case C'', respectively.
As shown in panel (a), all the cases successfully fit the observed Stokes $I$ profile.
Regarding the fitting of Stokes $Q$ and $U$ profiles shown in panels (b) and (c), we find similar synthetic profiles in all three cases.
This is due to the Van-Vleck ambiguity of the Hanle effect.
However, for the fitting of Stokes $V$ profile in panel (d), while ``case B'' and ``case C'' with green and blue lines show similar profiles that well fit the observation, ``case A'' with red line shows a significant difference from the observation.
The physical parameters at the pixel of interest obtained from the inversion for three cases are summarized in Table \ref{tab4}. 
These results suggest that there are two plausible solutions; $i.e.$, ``case B'' and ``case C''.
\\
~ In Figure \ref{fig421}, we show the spatial distribution of three inversion results of ``case A'', ``case B'', and ``case C''.
In panels (a), (e), and (i), we show the transverse component of the magnetic field in observer's frame for ``case A'', ``case B'', and ``case C'', respectively.
In panels (b), (f), and (j), we display the longitudinal component of the magnetic field in observer's frame for ``case A'', ``case B'', and ``case C'', respectively.
In panels (c), (g), and (k), the $\chi^{2}$ of Stokes $V$ of the ``case A'', ``case B'', and ``case C'' are shown, respectively.
Here we note that the $\chi^{2}$ is defined as $\chi^{2}=\Sigma_{n}|V_{\mathrm{obs}}(\lambda_{n})-V_{\mathrm{syn}}(\lambda_{n})|$, where $V_{\mathrm{obs}}$ and $V_{\mathrm{syn}}$ represent the observed and synthetic Stokes $V$ profiles, respectively.
In the three panels of the $\chi^{2}$ distributions, we can clearly find that the $\chi^{2}$ of Stokes $V$ for ``case A'' shows significantly higher values compared to the other two cases as expected from the synthetic Stokes $V$ profiles.
Thus, we can discard the ``case A'' solution.
In contrast, we find similar $\chi^{2}$ distributions of Stokes $V$ in ``case B'' and ``case C'', and the longitudinal components in observer's frame ($B_{\mathrm{l}}$) show similar distributions in these two cases (see panels (f) and (j)).
However, we find that the transverse component ($\bm{B_{\mathrm{t}}}$) of ``case B'' and ``case C'' are perpendicular to each other, $i.e.,$ these two cases are in the Van-Vleck ambiguity (see panels (e) and (i)).
To choose the proper solution, we compared the direction of $\bm{B_{\mathrm{t}}}$ with the threads of dark filament which is found in H$\alpha$ slit-jaw images of our observation and EUV images in $304$ $\mathrm{\AA}$  by the Atmospheric Imaging Assembly (AIA: \cite{Lemen2012}) on board the Solar Dynamics Observatory (SDO: \cite{Pesnell2012}).
In panels (h) and (d), we show the direction of threads of the filament with red and green arrows on the H$\alpha$ and 304 $\mathrm{\AA}$ images, respectively.
It is found that the directions of threads seen in H$\alpha$ and 304 $\mathrm{\AA}$ images are consistent and they fit the direction of $\bm{B_{\mathrm{t}}}$ of ``case B'' rather than that of ``case C''.
Since threads are likely aligned with the magnetic fields, we concluded that the best solution among the three cases for DF2 is the ``case B''.
\\ 
~ Regarding the $180$-degree ambiguity, $i.e.,$ the presence of two solutions with different azimuth angles in the local frame, one of them was already selected from signs of Stokes $V$ signal in HAZEL inversion by fitting the observed Stokes $V$ profiles.
We performed the same analysis for all the targets, and the results are summarized in Table \ref{tab5}.
Here we note that the obtained magnetic field is more or less uniform in a major part of the filament body and the values of the longitudinal component ($B_{\mathrm{l}}$), the field strength ($|\bm{B}|$), the inclination angle ($\theta$), the azimuth angle ($\phi$), and the optical thickness ($\tau$) shown in Table \ref{tab5} are the median values in the dark filament region.

\begin{threeparttable}[htb]
  \caption{Inversion results for three cases at the pixel of interest}
  \begin{center}
    \label{tab4}
    \begin{tabular}{lrrrrrrr}
      \hline
                 & $|\bm{B}|$ & $\theta$ & $\phi$ & $\tau$  & $v_{\mathrm{turb}}$ & $a$    & $ff$ \tnote{1}  \\
                 & [Gauss]     & [deg]    & [deg]  &         & [km/s]            &        &        \\
      \hline
      ``case A'' & $4.0$  & $24.1$  & $-11.7$ & $0.80$ & $5.0$ & $0.76$ & $1.0$  \\
      ``case B'' & $18.4$ & $99.7$  & $130.0$ & $0.80$ & $5.0$ & $0.74$ & $1.0$  \\
      ``case C'' & $14.0$ & $156.4$ & $156.8$ & $0.79$ & $5.0$ & $0.76$ & $1.0$  \\
      \hline
    \end{tabular}
    \begin{tablenotes}
      \item[1] Filling factor was set to 1 for all the cases.
    \end{tablenotes}
  \end{center}
\end{threeparttable}

\begin{figure}[htb]
  \begin{center}
    \includegraphics[bb= 0 0 1025 895, width=150mm]{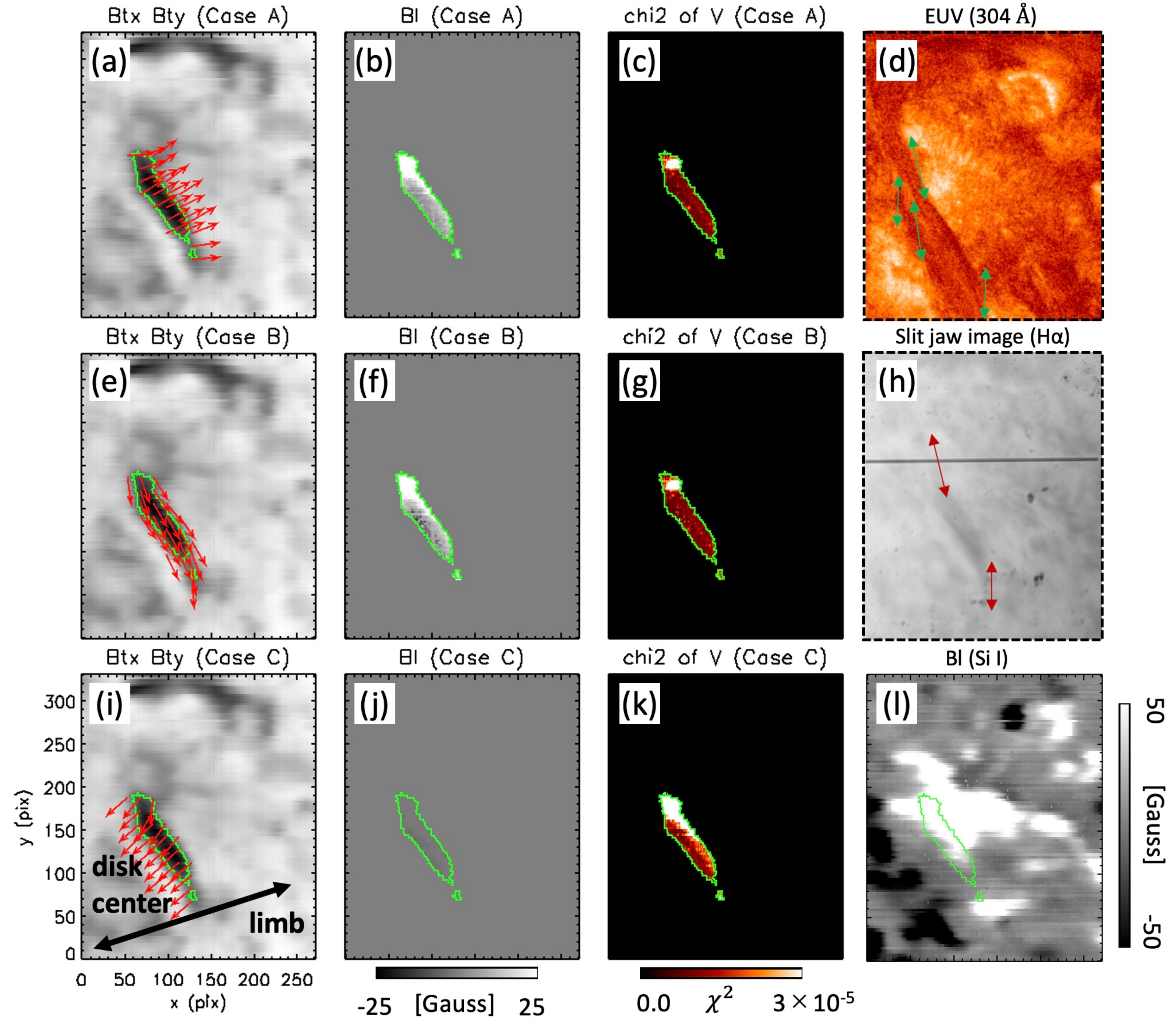}
  \end{center}
  \caption{Vector magnetic field for DF3. The formats of this figure is same as that of Figure \ref{fig421}.}\label{fig422}
\end{figure}

\subsection{Magnetic field configuration} \label{sec:43}
To clarify the type of magnetic field configuration of the filaments, we compared the direction of the transverse magnetic field ($\bm{B_{\mathrm{t}}}$) in the filaments and the direction of the global magnetic field expected from the photospheric magnetic field.
In Figure \ref{fig421} (l), we show the line-of-sight component of the photospheric magnetic field obtained from the Si I $10827$ $\mathrm{\AA}$ line.
As shown in panel (e), the direction of the transverse component of the magnetic field of the target filament is a top-to-bottom direction as a whole.
On the other hand, as shown in panel (l), in photosphere, the bottom-left side and the top-right side of the filament have positive and negative polarities, respectively, $i.e.$, directions of the global and the filament body magnetic field are opposite.
Thus, the magnetic field configuration of this target, DF2, can be interpreted as the reverse polarity.
\\
~ In Figure \ref{fig422}, we show an example of our target that has a normal polarity configuration, DF3.
By following the same process as DF2, we concluded that the ``case B'' result was the best solution for DF3.
In panel (e), we display the transverse component of the magnetic field in the filament body with red arrows.
In panel (l), the line-of-sight component of the photospheric magnetic field obtained from Si I $10827$ $\mathrm{\AA}$ is shown.
Taking into account both the transverse component of the filament's magnetic field and the photospheric magnetic field, we can find that the former is in the direction from top-left to bottom-right and the latter is in the direction from top to bottom in the field-of-view, respectively.
This means that this filament has a normal polarity configuration.
\\
~ Performing the same analysis for the other 6 targets, we found 7 out of 8 targets have the reverse polarity configuration (DF1, DF2, DF4, DF5, DF6, DF7, and DF8) and 1 out of 8 targets have the normal polarity configuration (DF3).
The results are summarized in the last column of Table \ref{tab5}.

\clearpage
\begin{threeparttable}[htb]
  \caption{Magnetic field parameters and magnetic field configuration of the targets ($\theta$ and $\phi$ in local frame)}
  \begin{center}
    \label{tab5}
    \begin{tabular}{lrrrrrc}
      \hline
      ID            & $|B_{\mathrm{l}}|$     & $|\bm{B}|$  & $\theta$  &  $\phi$  & $\tau$  &  Magnetic field \\
                    & [Gauss]              & [Gauss]     & [deg]     & [deg]    &         & configuration \tnote{1} \\
      \hline
      DF1 \tnote{2} & 24  &  35         &  63    &  52  & 0.63 & R          \\
      DF2           & 16  &  26         &  99    & 136  & 0.61 & R          \\
      DF3 \tnote{2} & 13  &  21         &  69    &  54  & 0.70 & N          \\
      DF4           & 2   &  8          & 105    & 105  & 0.76 & R          \\
      DF5           & 5   &  11         &  80    &  13  & 0.49 & R          \\
      DF6           & 0   &  17         &  88    &  89  & 0.58 & R          \\
      DF7 \tnote{2} & 1   &  15         &  80    &  87  & 0.57 & R          \\
      DF8 \tnote{3} & 1   &  11         &  76    &  94  & 0.38 & R          \\
     \hline
    \end{tabular}
    \begin{tablenotes}
      \item[1] ``N'' and ``R'' represent the normal and reverse polarity models, respectively. 
      \item[2] Targets locate above plage regions.
      \item[3] One footpoint of the filament is connected to an active region.
    \end{tablenotes}
  \end{center}
\end{threeparttable}


\section{Discussion \& Conclusion}\label{sec:dis}
As we presented in Table \ref{tab5}, we found that the field strength ($|\bm{B}|$) of our target filaments is in a range of $8-35$ Gauss. 
These results are consistent with the previous studies of QS filaments (see Table \ref{tab1}).
However, as \citet{DiazBaso2016} pointed out, the Stokes $V$ signals of dark filament regions can be created by the magnetic field in the photosphere below (see also \cite{DiazBaso2019a} and \cite{DiazBaso2019b}).
Therefore, since some of our targets are on plage region (DF1, DF3, and DF7) the magnetic field strength of these targets may be overestimated.
Regarding the inclination angles of the magnetic field in the local frame, they vary from $63-105^{\circ}$.
These results suggest that the magnetic field of dark filaments is more or less horizontal rather than vertical, which is consistent with previous work (e.g. \cite{TrujilloBueno2002}).
\\
~ As we presented in Table \ref{tab5}, we found that 7 out of 8 target DFs have reverse polarity configuration (DF1, DF2, DF4, DF5, DF6, DF7, and DF8) and 1 out of 8 targets have normal polarity configuration (DF3).
For the other 6 samples besides DF2 and DF3, we display the results in the same manner as Figures \ref{fig421} and \ref{fig422} in the Appendix.
According to the previous research, \citet{Bommier1998} performed a polarimetric observation in the He I D3 $5876$ $\mathrm{\AA}$ line for off-limb prominences and reported that 264 out of 296 QS prominences have reverse polarity in their magnetic field configuration.
By observing an off-limb prominence with He I $10830$ $\mathrm{\AA}$ line, \citet{Martinez2015} also reported that the magnetic field configuration of the prominence has the reverse polarity.
Our result of the magnetic field configuration of QS filaments is consistent with the result in previous studies.
So far most studies for determining the prominence field configuration as one of the normal or reverse  polarity types have been conducted with off-limb observations.
One exception is the work by \citet{Wang2020} who performed a spectropolarimetric observation in He I $10830$ $\mathrm{\AA}$ line for an on-disk QS filament and reported that the magnetic field configuration was reverse polarity.
Our study provides the first observation of multiple on-disk QS filaments and confirmed that the majority of the magnetic field configuration is the reverse polarity.

\begin{ack}
  We thank the staff members and students at Hida observatory for continuous support for our observations. 
  This work was supported by JSPS KAKENHI Grant Number JP21J14036.
  SDO is a mission of NASA's Living With a Star Program.
\end{ack}

\appendix
\section*{Magnetic field configuration for all the other targets}\label{sec:app}
\begin{figure}[htb]
  \begin{center}
    \includegraphics[bb= 0 0 1025 895, width=150mm]{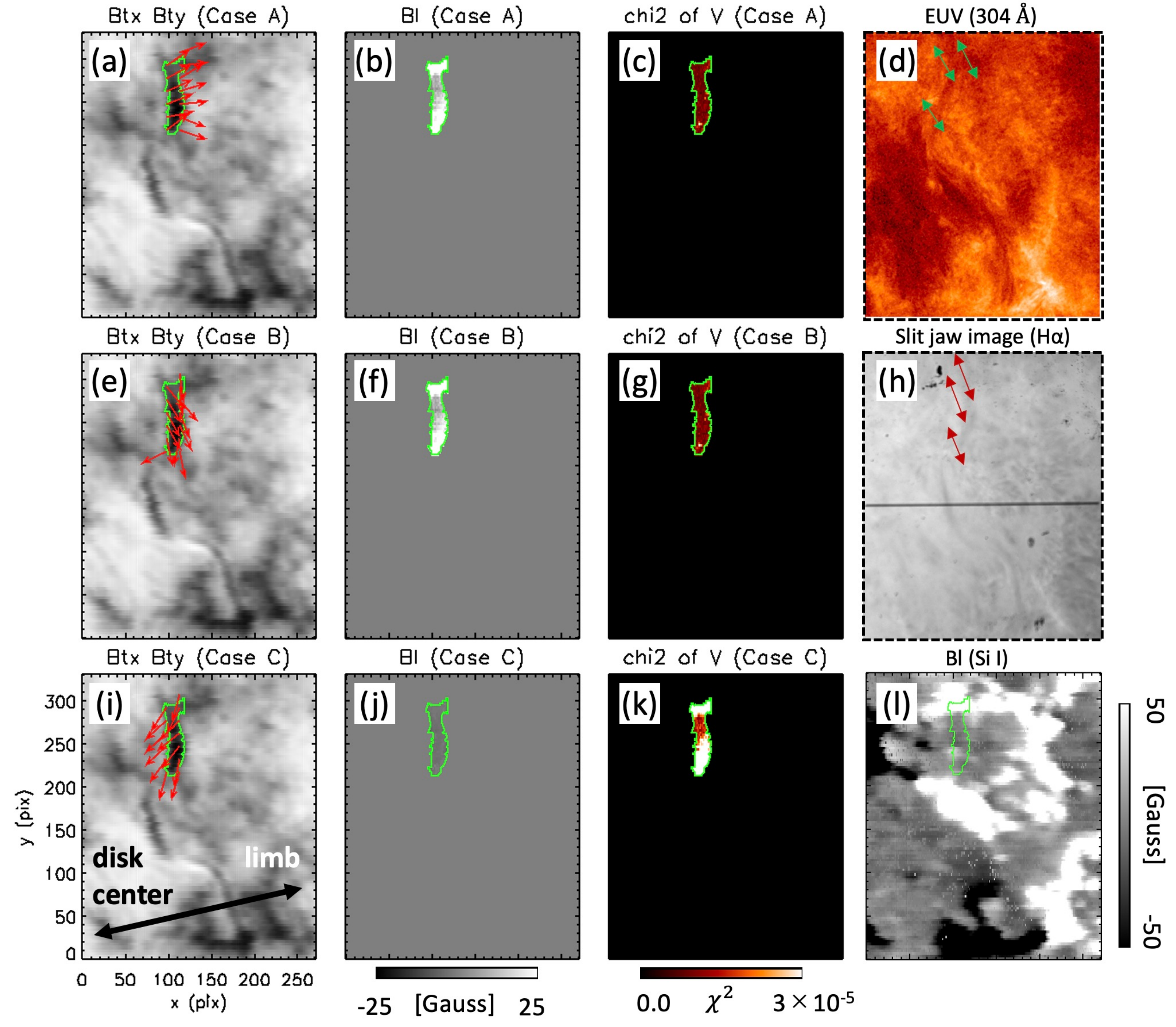}
  \end{center}
  \caption{Vector magnetic field for DF1. The formats of this figure is same as that of Figure \ref{fig421}.}\label{figa1}
\end{figure}

\begin{figure}[htb]
  \begin{center}
    \includegraphics[bb= 0 0 1120 895, width=150mm]{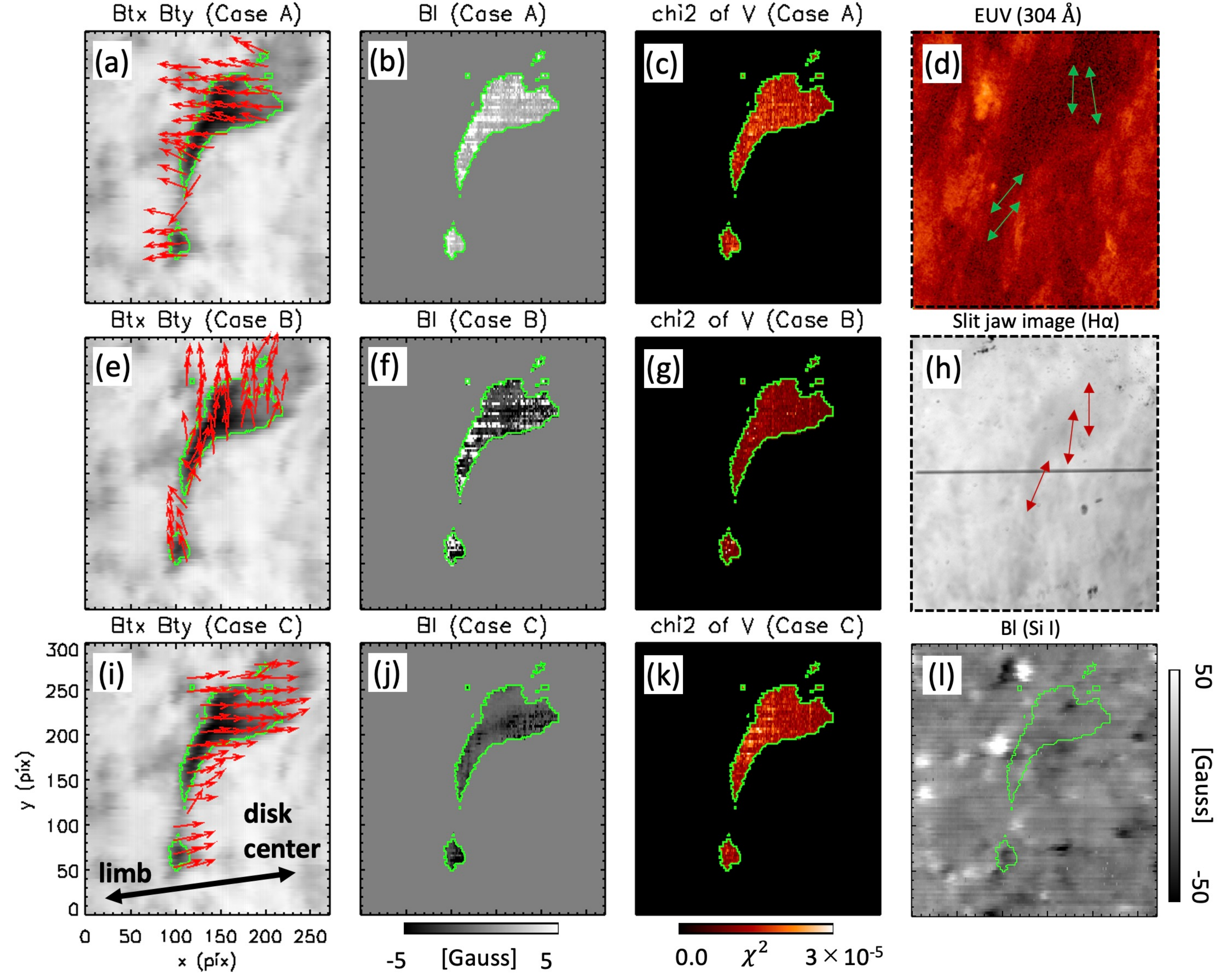}
  \end{center}
  \caption{Vector magnetic field for DF4. The formats of this figure is same as that of Figure \ref{fig421}. In these panels, we only show center part of the observation with $151''$ in $Y$ direction.}\label{figa2}
\end{figure}

\begin{figure}[htb]
  \begin{center}
    \includegraphics[bb= 0 0 1130 845, width=150mm]{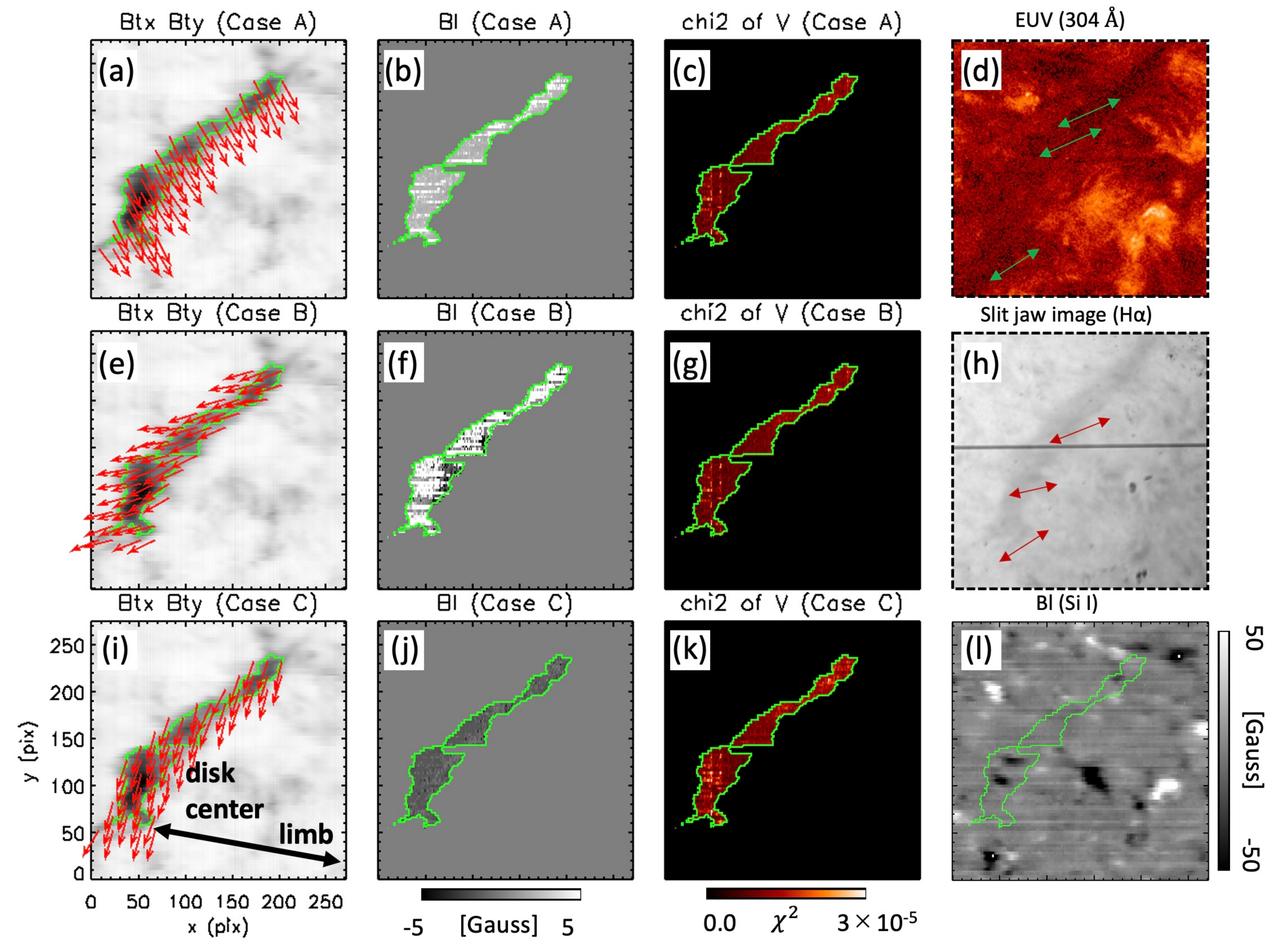}
  \end{center}
  \caption{Vector magnetic field for DF5. The formats of this figure is same as that of Figure \ref{fig421}.}\label{figa3}
\end{figure}

\begin{figure}[htb]
  \begin{center}
    \includegraphics[bb= 0 0 1130 840, width=150mm]{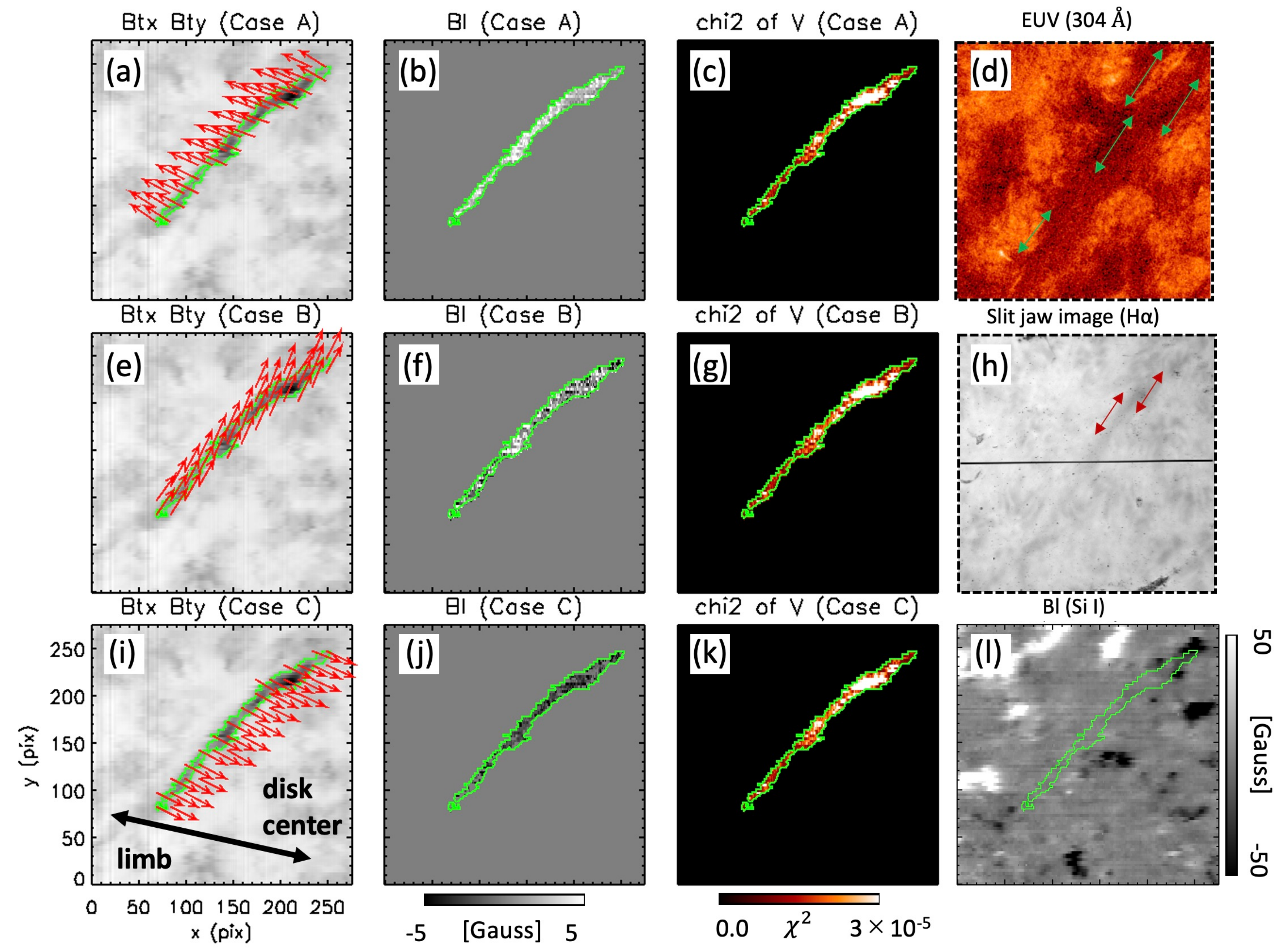}
  \end{center}
  \caption{Vector magnetic field for DF6. The formats of this figure is same as that of Figure \ref{fig421}.}\label{figa4}
\end{figure}

\begin{figure}[htb]
  \begin{center}
    \includegraphics[bb= 0 0 835 895, width=150mm]{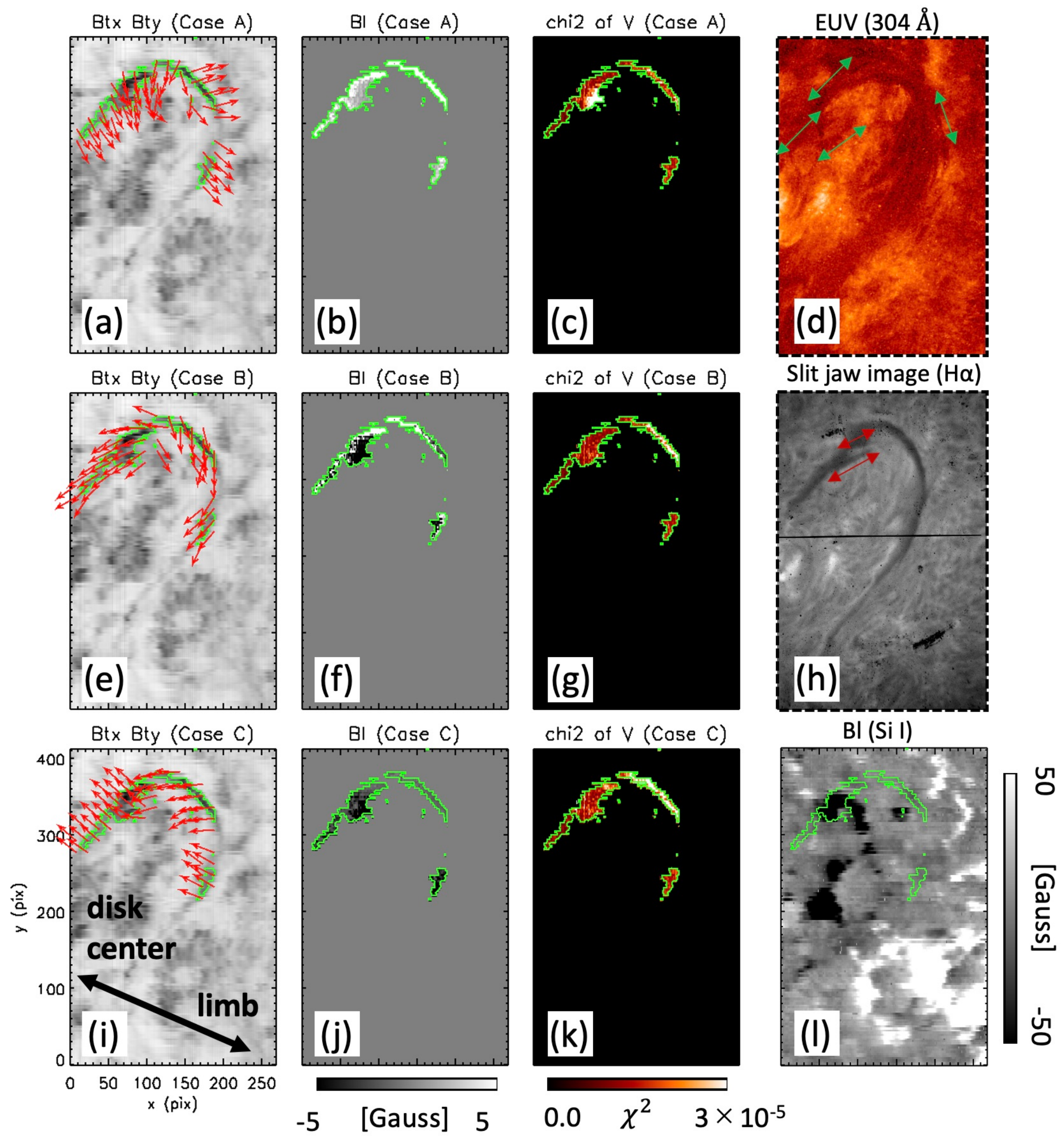}
  \end{center}
  \caption{Vector magnetic field for DF7. The formats of this figure is same as that of Figure \ref{fig421}.}\label{figa5}
\end{figure}

\begin{figure}[htb]
  \begin{center}
    \includegraphics[bb= 0 0 1090 845, width=150mm]{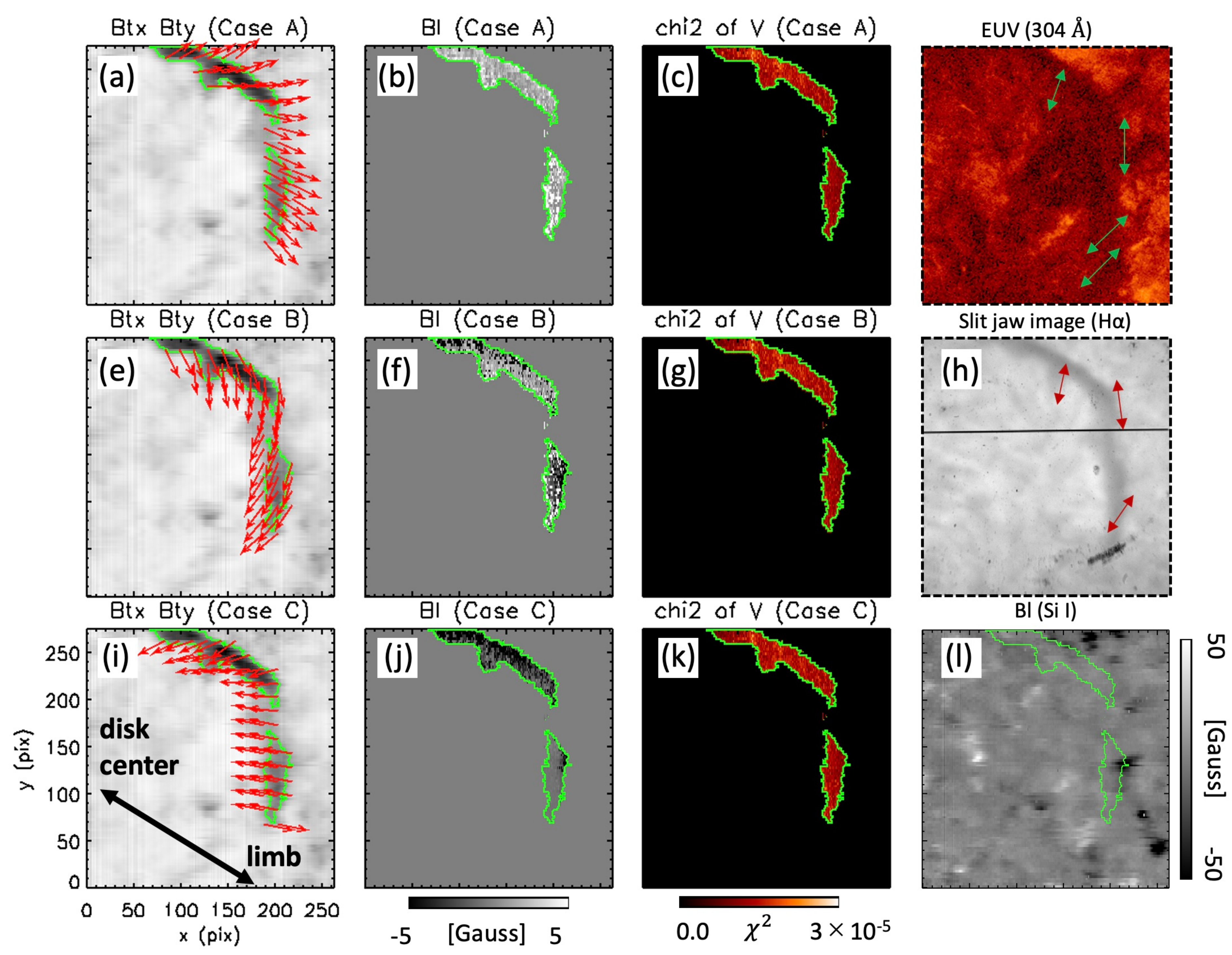}
  \end{center}
  \caption{Vector magnetic field for DF8. The formats of this figure is same as that of Figure \ref{fig421}.}\label{figa6}
\end{figure}

In Figures \ref{figa1}, \ref{figa2}, \ref{figa3}, \ref{figa4}, \ref{figa5}, and \ref{figa6}, we display the magnetic field configuration for DF1, DF4, DF5, DF6, DF7, and DF8, respectively.
Panels (a), (b), and (c) in each figure show transverse magnetic field vector in observer's frame with red arrows overplotted on Stokes I at the line center of He I 10830 $\mathrm{\AA}$, longitudinal magnetic field from He I $10830$ $\mathrm{\AA}$, and the line-of-sight component of the photospheric magnetic field from Si I $10827$ $\mathrm{\AA}$, respectively.
White- and black-colored regions in panels (b) and (c) are positive and negative polarities, respectively.
Green contour in all panels shows the border of the target dark filament.
Inversion results (median values in the filament) for these target DFs are listed in Table \ref{tab5}.
We note that we choose ``case B'' result as the proper solution for all the targets we display in this Appendix.

\end{document}